\documentclass[11pt,draftcls,onecolumn]{IEEEtran}
\IEEEoverridecommandlockouts

\usepackage{epsfig}
\usepackage{amsthm}
\usepackage{amsmath}
\usepackage{amssymb}
\usepackage{setspace}

\usepackage{psfrag}
\usepackage{url}
\usepackage{stfloats}
\usepackage{array}
\usepackage[version=3]{mhchem}
\usepackage{caption}
\usepackage{ulem}

\usepackage{amsfonts}

\usepackage{multirow}

\usepackage{dsfont}
\usepackage[latin1]{inputenc}
\usepackage{times}
\usepackage{color}
\usepackage{url}

\usepackage[mathscr]{euscript}
\usepackage{scalerel}
\usepackage{amssymb}

\usepackage{multibib}

\theoremstyle{definition}

\newtheorem*{example*}{Example}
\newtheorem*{remark*}{Remark}

\newcommand{\bre}{\begin{equation}}
\newcommand{\ere}{\end{equation}}
\newcommand{\be}\[
\newcommand{\ee}\]
\newcommand{\bra}{\begin{eqnarray}}
\newcommand{\era}{\end{eqnarray}}
\newcommand{\ba}{\begin{eqnarray*}}
\newcommand{\ea}{\end{eqnarray*}}
\newcommand{\bfg}{\begin{figure}[hbtp]}
\newcommand{\efg}{\end{figure}}
\newcommand{\bver}{\begin{verbatim}}
\newcommand{\ever}{\end{verbatim}}
\newcommand{\bit}{\begin{itemize}}
\newcommand{\eit}{\end{itemize}}
\newcommand{\ben}{\begin{enumerate}}
\newcommand{\een}{\end{enumerate}}

\newcommand{\bs}{{\bf s}}

\def\argmax{\mathop{\rm argmax}}

\newcommand{\comment}[1]{}

\def\etal{$et \,\,al.\,\,$}

\newlength{\tmpbigbar}

\newfont{\boldlarge}{msbm10 scaled 1100}
\newfont{\boldsmall}{msbm10 scaled 800}

\newcommand{\Expt}{\mbox{\boldlarge E}}

\newcommand{\eps}{\varepsilon}

\newcommand{\m}[1]{\mathcal{#1}}

\def\ll{<\hspace{-0.15cm}<}

\newtheorem{thm}{Theorem}
\newtheorem{rem}{Remark}
\newtheorem{cor}{Corollary}
\newtheorem{lem}{Lemma}
\newtheorem{dfn}{Definition}

\def\mb{\mathbf}

\def\bs{\boldsymbol}

\def\ss#1{{\sf #1}}

\def\vec#1{\mb{#1}}
\def\rvec#1{\bs{\uppercase{#1}}} 
\def\rvecr#1{\bs{\lowercase{#1}}} 

\def\mat#1{\mb{\uppercase{#1}}}

\def\EspOp{\ss{E}}

\newcommand{\Esp}[2][5]{%
  \ifcase#1
     \EspOp\{ #2 \}
     \or \EspOp \bigl\{ #2 \bigr\}
     \or \EspOp \Bigl\{ #2 \Bigr\}
     \or \EspOp \biggl\{ #2 \biggr\}
     \or \EspOp \Biggl\{ #2 \Biggr\}
  \else
     \EspOp \left\{ #2  \right\}
\fi}

\newcommand{\Earg}[3][5]{%
  \ifcase#1
     \EspOp_{#3} \{ #2 \}
     \or \EspOp_{#3} \bigl\{ #2 \bigr\}
     \or \EspOp_{#3} \Bigl\{ #2 \Bigr\}
     \or \EspOp_{#3} \biggl\{ #2 \biggr\}
     \or \EspOp_{#3} \Biggl\{ #2 \Biggr\}
  \else
     \EspOp_{#3} \left\{ #2  \right\}
\fi}

\newcommand{\CEsp}[3][5]{%
  \ifcase#1
     \EspOp\{ #2 \mid #3 \}
     \or \EspOp \bigl\{ #2 \bigm\vert #3 \bigr\}
     \or \EspOp \Bigl\{ #2 \Bigm\vert #3 \Bigr\}
     \or \EspOp \biggl\{ #2 \biggm\vert #3 \biggr\}
     \or \EspOp \Biggl\{ #2 \Biggm\vert #3 \Biggr\}
  \else
     \EspOp \left\{ #2  \,\middle\vert\, #3 \right\}
\fi}

\newcommand{\Diag}[2][5]{%
  \ifcase#1
     \mb{Diag}( #2 )
     \or \mb{Diag} \bigl( #2 \bigr)
     \or \mb{Diag} \Bigl( #2 \Bigr)
     \or \mb{Diag} \biggl( #2 \biggr)
     \or \mb{Diag} \Biggl( #2 \Biggr)
  \else
     \mb{Diag} \left( #2  \right)
\fi}

\newcommand{\diag}[2][5]{%
  \ifcase#1
     \mb{diag}( #2 )
     \or \mb{diag} \bigl( #2 \bigr)
     \or \mb{diag} \Bigl( #2 \Bigr)
     \or \mb{diag} \biggl( #2 \biggr)
     \or \mb{diag} \Biggl( #2 \Biggr)
  \else
     \mb{diag} \left( #2  \right)
\fi}

\def\log{\ss{log}}

\def\ie{i.e.}
\def\eg{e.g.}

\def\etal{et al.}

\newsavebox\myboxA
\newsavebox\myboxB
\newlength\mylenA

\newcommand*\xoverline[2][0.75]{%
    \sbox{\myboxA}{$\m@th#2$}%
    \setbox\myboxB\null
    \ht\myboxB=\ht\myboxA%
    \dp\myboxB=\dp\myboxA%
    \wd\myboxB=#1\wd\myboxA
    \sbox\myboxB{$\m@th\overline{\copy\myboxB}$}
    \setlength\mylenA{\the\wd\myboxA}
    \addtolength\mylenA{-\the\wd\myboxB}%
    \ifdim\wd\myboxB<\wd\myboxA%
       \rlap{\hskip 0.5\mylenA\usebox\myboxB}{\usebox\myboxA}%
    \else
        \hskip -0.5\mylenA\rlap{\usebox\myboxA}{\hskip 0.5\mylenA\usebox\myboxB}%
    \fi}

\def\fakeH{\rlap{\'{}\'{}}}

\newcommand{\gto}[1]{\to_{{\scriptscriptstyle \m{#1}} }}
\newcommand{\ngto}[1]{\not\to_{{\scriptscriptstyle \m{#1}} }}

\makeatletter
\def\ps@IEEEtitlepagestyle{%
\def\@oddfoot{\mycopyrightnotice}%
\def\@evenfoot{}%
}
\def\mycopyrightnotice{%
{\footnotesize 978-1-6654-4407-1/21/\$31.00~\copyright~2021 IEEE\hfill} 
\gdef\mycopyrightnotice{}
}

\begin{document}

\title{On lossy Compression of Directed Graphs}
\author{\IEEEauthorblockN{Ronit Bustin\footnote{The work was done while R. Bustin was a postdoc at Tel Aviv University.} and Ofer Shayevitz}
\thanks{This work has been supported the by the European Research Council, under grant agreement 639573.\newline The paper was presented in part at ISIT 2017. }\\
\IEEEauthorblockA{\scriptsize{Dept. of Electrical Engineering}\\
\scriptsize{Tel Aviv University, Israel}\\
\scriptsize{Email: ronit.bustin@gmail.com, ofersha@eng.tau.ac.il}} }

\maketitle
\normalem

\begin{abstract}
The method of types presented by Csisz\'ar and K\"{o}rner is a central tool used to develop and analyze the basic properties and constraints on sequences of data over finite alphabets. A central problem considered using these tools is that of data compression, and specifically lossy data compression. In this work we consider this very problem, however, instead of sequences of data we consider directed graphs. We show that given a more natural distortion measure, fitting the data structure of a directed graph, the method of types cannot be applied. The suggested distortion measure aims to preserves the local structure of a directed graph. We build on the recent work of Barvinok and extend the method of types to the two dimensional setting of directed graphs. We see that the extension is quite natural in many ways. Given this extension we provide a lower and upper bound on the rate-distortion problem of lossy compression given the suggested distortion measure.
\end{abstract}

\section{Introduction} \label{sec:intro}

Data compression is the task of representing a complex object using a small number of bits, while providing some guarantees on the quality of reconstruction. In its canonical form, this object typically comes in the shape of a sequence over some given alphabet, often assumed to be generated by some statistical model which is known either exactly or partially.  These types of problems and their many variations have been extensively studied in information theory, and the corresponding fundamental limits are generally well established (see e.g.~\cite{cover,Berger_book}). The situation is quite different however when it comes to compression of graphical objects, a topic of much contemporary interest that is far less explored and only scantly understood. Loosely speaking, the main distinction from the traditional paradigm, and also the main difficulty, lies in the fact that graphs are objects that encode {\em relations between elements}, rather than the elements themselves. The tools developed for the study of non-graphical objects generally fall short of addressing this issue.

The problem of graph compression has been extensively addressed in the literature over the last two decades. However, the vast majority of works have been dedicated either to algorithmic and experimental issues, see e.g. \cite{feder1995clique,suel2001compressing,adler2001towards,boldi2004webgraph,chierichetti2009compressing,apostolico2009graph,hayashida2010comparing,nourbakhsh2015matrix,mohri2015automata} and references therein, or to entropy calculations for certain families of (labeled) random graphs, see e.g.~\cite{aldous2014entropy,mohri2015automata}. Attempts to study the fundamental problems endemic to graph compression in greater information-theoretic rigor are only very recent. Choi and Szpankowski~\cite{choi2012compression} considered {\em lossless} compression of {\em unlabeled Erd\fakeH{o}s-R\'enyi random graphs}, where the graph needs to be perfectly reconstructed up to isomorphism. They computed what they called the {\em structural entropy} of the graph, which is the entropy of the isomorphism equivalence classes, a quantity that clearly constitutes the fundamental compression limit in this setup. They also provided a polynomial-time algorithm that approaches the structural entropy. Anantharam and Delgosha~\cite{delgosha2017universal} attacked the general problem of (lossless) {\em universal} compression of (vertex- and edge-labeled) graphs under no statistical assumptions. To that end, they defined a suitable limit (in the local weak convergence sense) for sequences of growing graphs, accompanied by a corresponding notion of entropy for the limiting object. They showed that it is possible to losslessly compress sequences that converge in this sense, with a rate asymptotically approaching the limit entropy, without knowing the limit in advance.

Our approach differs from these two aforementioned works in several important ways. Unlike~\cite{choi2012compression}, we consider {\em lossy} compression of {\em labeled} graphs (where vertex identity matters). For labeled graphs, the lossless compression problem with a known distribution reduces to the traditional setup and is less interesting. For lossy compression however, where one wished to reconstruct the graph but allows some limited inaccuracy in the {\em local structure} around each vertex, the problem diverges from the traditional setup and becomes very challenging. To the best of our knowledge, this type of problem has not been studied before in the literature. Similarly to~\cite{delgosha2017universal} we too consider {\em universal} compression of graphs, and our approach bears some similarities to their ``depth $h$ empirical distribution''. However, unlike~\cite{delgosha2017universal} where results are asymptotic and require convergence, and where no underlying {structure} is assumed, we study universality for {\em fixed-sized graphs} with structural constraints.

Below we introduce a systematic approach that is inspired by, and significantly expands on, what is arguably the most successful analysis tool in traditional data compression over finite alphabets, i.e., the so-called {\em method-of types} introduced by Csisz\'ar and K\"{o}rner~\cite{csiszar_korner}. Roughly speaking, the method-of-types is a combinatorial tool that partitions the data space into equivalence classes, called {\em type classes}, that share common empirical statistics. For example, the (first-order) types associated with length-$n$ binary sequences are the Hamming spheres, i.e., two sequences are of the same type if they have the same fraction of '$1$'s. Analyzing quantities such as the sizes of type classes and the probability that a random sequence would land in a certain type, has proved immensely instrumental in obtaining the fundamental limits in various distinct data compression problems, as well as in other fields such as channel coding, large deviations, and discrete mathematics. 

The aim of this work is twofold, first to propose a graph type theory, and then show its applicability on lossy graph compression. The suggested graph type theory follows the ideas presented in the \emph{method of types} are can be extended and applied similarly, for example to hypothesis testing questions on graphs.
Our exposition of the various elements involved is problem-driven, meaning, we begin by discussing the problem of lossy graph compression and the difficulties it entails. We then propose an alternative distortion measure which better preserves the local structure of a graph. From this proposed distortion measure we define our lossy compression problem. Solving this problem requires new tools and results, which follow and extend upon the classical {method of types}. The remainder of the work focuses on this extensions and finally the bounds it can provide on the presented lossy compression problem.

\subsection{Paper Structure} \label{ssec:structure}
The remainder of this section contains the notation used in the paper and the main definition used throughout, which is that of the edge type. Section \ref{sec:CompressionProblemSetting} depicts the problem of lossy compression of graphs, suggesting a new distortion measure and defines the rate-distortion problem. A summary of known and relevant results is provided in Section \ref{sec:KnownResults}, specifically depicting the important contribution of Barvinok, on which this paper builds. In Section \ref{sec:MethodOfTypesExtended} the {method of types} is extended from one dimensional vector sequences to the two dimensional structure of directed graphs. The conditional case of this extension is considered in Section \ref{sec:ConditionalEdgeTypes}. We return to the rate-distortion problem using these tools in Section \ref{sec:RateDistortionGraphs}. Finally, we conclude the paper with a short summary in Section \ref{sec:summary}.


\subsection{Notation} \label{ssec:notation}
All graphs considered here are {\em directed graphs} over the same vertex set $[n] = \{1,2,\ldots,n\}$, unless otherwise stated. {These include graphs with self loops, meaning edges from a vertex to itself.} Most definitions and claims hold for undirected graphs with minor modifications. We write $i \gto{G} j$ to denote that a directed edge from $i$ to $j$ exists in the graph $\m{G}$, and $i \ngto{G} j$ otherwise. When there is no confusion, we will sometimes drop the subscript indicating the graph. A graph $\m{G}$ can be equivalently represented by its $n\times n$ binary adjacency matrix $\mat{A}_{\m{G}}$, where $\{\mat{A}_{\m{G}}\}_{ij}=\mathds{1}(i\gto{G} j)$. We write $\m{G}\oplus  \m{H}$ (resp. $\m{G}\wedge \m{H}$) to denote a graph whose adjacency matrix is obtained by a cell-wise XOR (resp. AND) of $\mat{A}_{\m{G}}$ and $\mat{A}_{\m{H}}$. We also consider {\em random graphs}, and specifically write $\m{F}\sim \{p_{ij}\}$ to mean that $i\gto{F} j$ with probability $p_{ij}$ independently over the edges. \\
The paper contains many new concepts and definitions. We have summarized these in a table available in Appendix \ref{appendix:TableOfNotations}.

\subsection{Main Definition - Edge Type} \label{ssec:mainDefinition}
Before proceeding to the exposition of the problem setting and an overview of central and relevant results from the literature, we want to focus on a central definition which holds the essence and main ideas leading to the analysis in this work. We will later enhance and generalize the definition somewhat, but the core is given in its basic form.
\begin{dfn} \label{dfn:edgeType}
Let $\m{G}$ be a directed graph over $[n]$. Let $\vec{r}_{\m{G}}$ and $\vec{c}_{\m{G}}$ be the vectors recording the number of outgoing and ingoing edges from/to each of the vertices, respectively. Namely,
\begin{align}
  \vec{r}_{\m{G}}(i) \triangleq \sum_{j\in[n]}\mathds{1}(i\gto{G} j), \qquad \vec{c}_{\m{G}}(i) &\triangleq \sum_{j\in[n]}\mathds{1}(j\gto{G} i).
\end{align}
Note that the vectors $\vec{r}_{\m{G}}$ and $\vec{c}_{\m{G}}$ are simply the row and column sums of the adjacency matrix $\mat{A}_{\m{G}}$. We define the {\em edge-type} of $\m{G}$ to be $T_{\m{G}}\triangleq (\vec{r}_{\m{G}},\vec{c}_{\m{G}})$. The set of all graphs with the same edge-type is called an {\em edge-type class}. We denote the edge-type class of the directed graph $\m{G}$ as $\rvec{T}_{\m{G}}$, or as $\rvec{T}( \vec{r}, \vec{c})$ when the class is defined directly by the outgoing and ingoing degree.
\end{dfn}
The above definition allows us to connect graphs to the method-of-type, and thus examine the graph compression problem similarly to the compression of strings (vectors) which builds on the method-of-type.

An important extension of the above definition produces a subset of the \emph{edge-type class} by further restricting the set of graphs
\begin{dfn} \label{dfn:RestrictedEdgeType}
The \emph{restricted edge-type class} $\rvec{T}( \vec{r}, \vec{c}, \m{W})$ is the set of directed graphs $\m{G}$ such that
\begin{align}
\m{G} \in \rvec{T}( \vec{r}, \vec{c}) \textrm{ and } \m{G} = \m{G} \wedge \m{W}.
\end{align}
{We also denote it as $\rvec{T}( \m{G}, \m{W})$, when it is directly defined by a graph $\m{G}$ and a restriction graph $\m{W}$.}
The corresponding \emph{restricted edge-type} is denoted by either $T = (\vec{r}, \vec{c}, \m{W})$ {or $T = (\m{G}, \m{W})$}.
\end{dfn}

Note that Definition \ref{dfn:RestrictedEdgeType} generalizes Definition \ref{dfn:edgeType} when $\m{W}$ is a complete digraph over $[n]$ (every pair of vertices is connected by a pair of edged, one in each direction). As such, in the sequel we will refer to he \emph{restricted edge-type class} simply as an \emph{edge-type class}, and to the \emph{restricted edge-type} as an \emph{edge-type}.

\section{Compression - Problem Setting} \label{sec:CompressionProblemSetting}
The problem of \emph{losslessly} compressing a random directed graph with a known distribution is straightforward in principle. In this case, the structure of the graph does not really come into play; one can simply think of the adjacency matrix as a random vector, and apply (say) arithmetic coding to compress it essentially to its entropy. This is true for any general distribution on the graph, and is particularly simple when the edges are chosen independently, in which case there is no need to compute (possibly difficult) conditional distributions.

The case of \emph{lossy} compression of random directed graphs can be very different, even when the distribution is known. Of course, if the distortion measure is defined additively over the edges, then the problem can again be translated into the traditional vector formalism. However, such a distortion measure makes little sense for graphs. To see why, suppose we would like to compress the graph such that the overall Hamming distortion over the edges is small, i.e., not too many edges ``appear'' or ``disappear'' when reconstructing the graph. This may sound reasonable, but note that under this criterion one is allowed to concentrate most, or even all of the distortion on a small number of vertices, and completely corrupt their neighborhoods. It is thus more reasonable to require that the distortion at {\em every vertex} is small. This leads us to consider the following distortion function on pairs of graphs:
\begin{align}\label{eq:distortion_edge_type}
  d(\m{G},\m{H}) \triangleq { \frac{1}{n}} \max\left\{\|\vec{r}_{\m{G}\oplus \m{H}}\|_\infty \,,\, \|\vec{c}_{\m{G}\oplus \m{H}}\|_\infty\right\}.
\end{align}
Clearly, if $\m{H}$ is used as a reconstruction for $\m{G}$, then the fraction of errors we make in the reconstruction of each vertex's immediate neighborhood is at most $d(\m{G},\m{H})$. This distortion measure takes into account the {\em local structure} of the graph, and cannot be directly handled using traditional lossy compression tools.

We define two types of rate-distortion functions associated with the above graph distortion measure.
The first is a combinatorial one:
\begin{dfn}[A combinatorial graph rate-distortion function] \label{dfn:CombinatorialRateDistortion}
\begin{align}
  & R_n(d) \triangleq \min_{\mathfrak{F}} \frac{\log |\mathfrak{F}|}{n^2} \nonumber \\
  & \text{s.t.}\qquad  \max_{\m{G}}\min_{\m{H}\in \mathfrak{F}} d(\m{G}, \m{H}) \leq d,
\end{align}
where the minimization is taken over subsets $\mathfrak{F}$ of graphs on $[n]$, and the maximization is over $\m{G}$, \ie, over all directed graphs on $[n]$. Namely, $0\leq R_n(d)\leq 1$ is the smallest rate (in bits per potential edge) that guarantees a local distortion of at most $d$ at each vertex, for any graph on $[n]$.

{The definition can be further generalized when we allow to consider a subset of all graphs on $[n]$ (as apposed to all graphs on $[n]$). We can denote such a subset as $S$. In the above the outer maximization will be over all graphs $\m{G} \in S$. An example for such an $S$ is the set of graphs on $[n]$ complying with a specific restriction graph $\m{W}$, or a specific edge-type class $\rvec{T}(\vec{r}, \vec{c}, \m{W})$.}
\end{dfn}

The second definition is probabilistic.
\begin{dfn}[A probabilistic graph rate-distortion function] \label{dfn:ProbabilisticRateDistortion}
For a random graph $\m{F}\sim\{p_{ij}\}$, let
\begin{align}\label{eq:prob_rd}
  & R_n^{\m{F}}(d,\eps) \triangleq \min_{\mathfrak{F}} \frac{\log |\mathfrak{F}|}{n^2}\nonumber \\
  & \text{s.t.}\qquad  \Pr\left(\min_{\m{H}\in \mathfrak{F}} d(\m{F}, \m{H}) > d\right) \leq  \eps.
\end{align}
Namely, $R_n^{\m{F}}(d,\eps)$ is the smallest rate (in bits per potential edge) that guarantees a local distortion of at most $d$ at each vertex with probability at least $1-\eps$, for the random graph $\m{F}$.
\end{dfn}
{In this work we consider the universal setting, meaning we consider a set of graphs without any assumption on the generating source of these graphs. As such, our goal is to provide insight to the combinatorial graph rate-distortion function, given in Definition \ref{dfn:CombinatorialRateDistortion}. We will do so by utilizing the probabilistic definition given in Definition \ref{dfn:ProbabilisticRateDistortion} and specifically the connection (see proof of \cite[Theorem 2.4.2]{csiszar_korner}):
\begin{align} \label{eq:connectionCombinatorialProbabilisticRateDistortion}
R_n(d) \geq \max_{\m{F}} \lim_{\epsilon \to 0} R_n^{\m{F}}(d,\epsilon)
\end{align}
where we assume $R_n^{\m{F}}(d,\epsilon)$ is well defined for all $\epsilon \in (0,1)$ and the maximization is over all random graphs that vanish outside the support of the relevant set $S$ of graphs, for which $R_n(d)$ is defined.}

\section{Relevant Results from the Literature} \label{sec:KnownResults}
Edge-types have been investigated since the 1950's, see e.g. Brualdi \cite{BrualdiSurvay,BrualdiBook2006} for an excellent survey. The vast majority of previous works have been written, however, in terms of binary matrices rather than graphs. This is, of course, an equivalent viewpoint, but it is more pertinent for our purposes to look at things from the graph perspective. Below we therefore recast some known results in the language of edge-types.

Surely both $\vec{r}$ and $\vec{c}$ may take one of $( n + 1)^n$ values, however not all pairs $(\vec{r},\vec{c})$ define a non-empty edge-type. Ryser \cite{Ryser1957} and Gale \cite{Gale1957} have independently characterized the set of all possible edge-types, i.e., gave a necessary and sufficient condition on a pair $(\vec{r},\vec{c})$ such that the associated edge-type class is nonempty. Their result is the following:
{
\begin{thm}{[\cite[Theorem 1.1]{Ryser1960}, \cite{Gale1957}]} \label{thm:GaleRyser}
Let $\vec{r}$ and $\vec{c}$ be two vectors whose components are nonnegative integers. {For a given $\vec{r}$ one can construct a matrix $\bar{\mat{A}}$, of size $n \times n$, by having each row be constructed as follows} {One can construct a matrix $\bar{\mat{A}}$, of size $n \times n$, complying with the row sum vector $\vec{r}$, by having each row constructed as follows}
\begin{align}
( 1, 1, \ldots, 1, 0, \ldots, 0)
\end{align}
meaning first all $1$'s and then all $0$'s. Such a matrix is sometimes referred to in the literature as a \emph{maximal} matrix. The column sum vector of $\bar{\mat{A}}$ is denoted by $\bar{\vec{c}}$. A necessary and sufficient condition in order for the edge-type class $T (\vec{r},\vec{c})$ to be non-empty is that
\begin{align}
\vec{c} \prec \bar{\vec{c}}
\end{align}
meaning that $\bar{\vec{c}}$ majorizes over $\vec{c}$, or explicitly that
\begin{align}
\sum_{i = 1}^k \vec{c}^{\downarrow}(i) \leq \sum_{i = 1}^k \bar{\vec{c}}^{\downarrow}(i), \quad \forall k \in [n-1],
\end{align}
and
\begin{align}
\sum_{i = 1}^n \vec{c}^{\downarrow}(i) = \sum_{i = 1}^n \bar{\vec{c}}^{\downarrow}(i),
\end{align}
where for any real-values vector $\vec{a}$, $\vec{a}^{\downarrow}$ is a vector with the same components, but sorted in descending order.
\end{thm}
}

In \cite{Ryser1957} Ryser also termed the concept of an \emph{interchange} as a transformation of the elements of the adjacency matrix $\mat{A}_{\m{G}}$ of a given graph $\m{G}$, that changes a specific sub-matrix of the following structure
\begin{align}
\mat{A}_1 = \left[ \begin{array}{cc} 1 & 0 \\ 0 & 1 \end{array} \right]
\end{align}
into
\begin{align}
\mat{A}_2 = \left[ \begin{array}{cc} 0 & 1 \\ 1 & 0 \end{array} \right]
\end{align}
or vice-versa (a sub-matrix $\mat{A}_2$ into a sub-matrix $\mat{A}_1$), and leaves all other elements of $\mat{A}_{\m{G}}$ unaltered. It is simple to see that an \emph{interchange} results in a directed graph of the same edge-type class as $\m{G}$, \ie, $T_{\m{G}}$. In \cite{Ryser1957} Ryser has shown the following
\begin{thm}{\cite[Theorem 3.1]{Ryser1957}} \label{thm:RyserInterchangeConncetivity}
Let $\m{G}$ and $\m{H}$ be two directed graphs over $[n]$ with the same edge-types. Then $\m{G}$ is transferable into $\m{H}$ by a finite number of interchanges.
\end{thm}
Given the definition of an \emph{interchange} and the above result of Ryser, it is natural to define an \emph{interchange graph} as done by Ryser \cite{Ryser1957}:
\begin{dfn} \label{dfn:interchangeGraph}
An \emph{interchange graph} of an edge-type class $\rvec{T} (\vec{r}, \vec{c})$ is a non-directed graph whose vertices are all {matrices} {graphs} in $T (\vec{r}, \vec{c})$. Two vertices (directed graphs) are adjacent to one another if one can be produced from the other by a single \emph{interchange}.
\end{dfn}
Due to the above result by Ryser in Theorem \ref{thm:RyserInterchangeConncetivity} we have that the interchange graph is fully connected. The \emph{interchange graph} has been investigated and many additional results are available. There are however many open questions \cite{Qian1999,BrualdiBook2006}.

Ryser \cite{Ryser1960} has also introduced an interesting entity called the {\em structure matrix} of a directed graph. {A structure matrix of a directed graph over $[n]$ is} an integer-valued {$(n+1) \times (n+1)$} matrix that {is constructed directly from the vectors $\vec{r}$ and $\vec{c}$ and gives important insights to the structure of the graphs within the edge-type class. More specifically, the structure matrix} characterizes all the {\em invariant edges} of an edge-type, i.e., those edges that are guaranteed to always exist (or to never exist) in all the graphs within the edge-type class. {Note that the structure matrix contains the indices from the set $\{0, 1,2 \ldots, n\}$ for both rows and columns.}
{There are two issues with structure matrices that we need to mitigate:
\begin{itemize}
\item Structure matrices are constructed for normalized edge-types, which are edge-types in which both $\vec{r}$ and $\vec{c}$ are non-increasing vectors.
\item Structure matrices do not take into account restriction graphs, $\m{W}$.
\end{itemize}}

{We begin by providing the constructive definition for the structure matrix of a normalized edge-type:}
{\begin{dfn}[\cite{BrualdiSurvay}] \label{dfn:structureMatrix}
Let $\m{G}$ belong to a normalized class $\rvec{T}(\vec{r}, \vec{c})$ (this is an edge-type class with no restriction graph, or alternatively, $\m{W}$ is the fully connected graph). {We examine its adjacency matrix $\mat{A}_{\m{G}}$. For each location $(e,f)$ in the matrix we can denote it as follows:}
{For each pair $(e,f)$, $0 \leq e,f \leq n$, we write the adjacency matrix $\mat{A}_{\m{G}}$ in a block form as follows:}
\begin{align}
\mat{A}_{\m{G}} = \left[ \begin{array}{cc} \mat{W} & \mat{X} \\ \mat{Y} & \mat{Z} \end{array} \right],
\end{align}
where $\mat{W}$ is of size $e \times f$ {($0 \leq e \leq n, 0 \leq f \leq m$)}{. Now the} {and define the $(e,f)$ coordinate in the} structure matrix {$\mathbb{T}_{\m{G}}$ as:} {is constructed as follows:}
\begin{align}
\{\mathbb{T}_{\m{G}}\}_{ef} = N_0 ( \mat{W}) + N_1( \mat{Z})
\end{align}
where $N_0( \cdot)$ denotes the number of zeros in a matrix, and $N_1( \cdot )$ denotes the number of ones.
An easy calculation shows that
\begin{align}
\{\mathbb{T}_{\m{G}}\}_{ef} = e f + \sum_{i > e} \vec{r}(i) - \sum_{j \leq f} \vec{c}(j)
\end{align}
from which it is more evident that the \emph{structure matrix} is a property of the normalized edge-type class.
\end{dfn}}

{We now provide the definitions of invariant set and invariant position, as well as the definition of components which are central to the understanding of the importance of the structure matrix in extracting the properties of the edge-type class.}

\begin{dfn}[\cite{BrualdiSurvay}] \label{dfn:invariatSetinvariantPosition}
{For $K \subseteq [n]$ and $L \subseteq [n]$ we say that $K \times L$ is an invariant set for $\rvec{T} (\vec{r}, \vec{c})$ if for all $\m{G}$ and $\m{H}$ in $\rvec{T} (\vec{r}, \vec{c})$ we have that
\begin{align}
N_1 \left( \mat{A}_{\m{G}}[K,L] \right) = N_1 \left( \mat{A}_{\m{H}}[K,L] \right)
\end{align}
where $N_1( \cdot )$ returns the number of ones in the (sub)matrix. }

{
An invariant position is an invariant set of cardinality one. An invariant position is either an invariant 1-position, meaning the edge always exists in all graphs of the class, or an invariant 0-position, meaning the edge never exists in all graphs of the class.}
\end{dfn}

{
\begin{dfn}[\cite{BrualdiSurvay}] \label{dfn:components}
Suppose $I_1, \ldots, I_p$ are the minimal non-empty subsets of $[n]$ such that for $i=1,\ldots, p$ the column sum vector of $\mat{A}_{\m{G}} [ I_i, [n]]$ is the same for all $\m{G} \in \rvec{T}(\vec{r}, \vec{c})$. Suppose $J_1, \ldots, J_q$ are the minimal non-empty subsets of $[n]$ such that for $j=1, \ldots, q$ the row sum vector of $\mat{A}_{\m{G}} [ [n], J_j]$ is the same for all $\m{G} \in \rvec{T}(\vec{r}, \vec{c})$. Then, $I_1, \ldots, I_p$ and $J_1, \ldots, J_q$ are partitions of $[n]$. $I_1, \ldots, I_p$ are the row components and $J_1, \ldots, J_q$ are the column components and there exists nonnegative vectors $\vec{r}_1, \ldots, \vec{r}_p$ and $\vec{c}_1, \ldots, \vec{c}_q$ such that for any $\m{G} \in \rvec{T}(\vec{r}, \vec{c})$ the  sub-matrices of the adjacency matrix $\mat{A}_{\m{G}} [ I_i, J_j]$ will have a sum rows of $\vec{r}_i$ and sum columns of $\vec{c}_j$.
The sets $I_i \times J_j$ are called the components of $\rvec{T}(\vec{r}, \vec{c})$.
\end{dfn}
}

{The next theorem, due to Haber \cite{BrualdiSurvay}, is an improvement of a theorem by Ryser \cite{Ryser1957}.}
{
\begin{thm}[Theorem 5.2 \cite{BrualdiSurvay}] \label{thm:HuberRyser_invariantPosition_structureMatrix}
Assume $\rvec{T}(\vec{r}, \vec{c})$ is a normalized edge-type class. For all $\m{G} \in \rvec{T}(\vec{r}, \vec{c})$ the edge $i \gto{G} j$ exists for $i \in [e]$ and $j \in [f]$ (an invariant 1-position) while $i \ngto{G} j$ for $e < i \leq n$ and $f < j \leq n$ (an invariant 0-position) if and only if
\begin{align}
\left\{ \mathbb{T}_{\vec{r}, \vec{c}} \right\}_{e,f} = 0.
\end{align}
\end{thm}
}

{The above definitions and result come to show that an edge-type class has basic properties that can be extracted from the structure matrix. Specifically, the graphs within the edge-type class can all be similarly partitioned into components. These components are either trivial components in which all edges are invariant (either 1-position or 0-position), or nontrivial components. The nontrivial components contain no invariant edges, in other words, for every edge in a nontrivial component we may find a graph within the edge-type class which contains that edge and a graph which does not contain that edge. Note that we provided only a glimpse into the topic of structure matrices, for a more thorough review of this matrix, its construction and properties, the reader is referred to \cite{BrualdiSurvay} and \cite{BrualdiBook2006}.}

{A specific example of constructing a structure matrix and extracting from it the invariant edges is given in Appendix \ref{appendix:exampleStructureMatrix}. This example is taken from \cite{BrualdiSurvay} and is given here for completeness.}

{As will be evident in the sequel, identifying the invariant edges can become very useful. As an immediate example we can use them to define a necessary condition for a restricted edge-type class to be non-empty, as an extension of Theorem \ref{thm:GaleRyser}:
\begin{lem} \label{lem:requirementOnRestrictionGraph}
For a given pair $T = (\vec{r}, \vec{c})$, any restriction graph $\m{W}$ matched to this pair must comply with
\begin{align}
\m{W}_{T, invariant-1-position} \wedge \m{W} = \m{W}_{T, invariant-1-position}
\end{align}
as a necessary condition for $\rvec{T}(\vec{r}, \vec{c}, \m{W})$ to be non-empty.
$\m{W}_{T, invariant-1-position}$ is a graph containing only the invariant 1-position edges of the edge-type class $\rvec{T}(\vec{r},\vec{c})$.
\end{lem}
In other words, necessary conditions to have a non-empty restricted edge-type include both those in Theorem \ref{thm:GaleRyser} on the pair $(\vec{r},\vec{c})$ as well as the above condition on the restriction graph.
}

{Having acknowledges the importance of identifying the invariant edges we recall that two issues were raised that require our concern.}
%
{As mentioned above the structure matrix is defined for normalized edge-type classes. With no loss of generality we can rename the vertices such that $\vec{r}$ is normalized (ordered in non-increasing order). Still the resulting $\vec{c}$ is not necessarily normalized. Let us denote as $\vec{c}^{\downarrow}$ the non-increasing sorted $\vec{c}$. The relation between $\vec{c}^{\downarrow}$ and $\vec{c}$ defines a permutation on the columns of the corresponding adjacency matrices of the graphs in the edge-type class. We can denote the reverse {permutation} as the function $R(\cdot)$. Meaning, for a given vertex $j$ with $\vec{c}^{\downarrow}(j)$ ingoing degree, the actual vertex required to have that {value of} ingoing degree is $R(j)$.}

{
We can construct the structure matrix for the normalized {edge-type class} $\rvec{T}(\vec{r}, \vec{c}^{\downarrow})$. This matrix points out which edges are invariant. However if edge $i \gto{} j$ is invariant according to this structure matrix, we actually conclude that the invariant edge is $i \gto{} R(j)$. In other words, from the structure matrix for the normalized edge-type class and the function $R(\cdot)$ we can identify the invariant positions.}

{
The second issue is more complex. The observations we will see in the sequel apply to nontrivial components. However, the structure matrix allows us to identify such components only in the absence of additional restrictions (such as those in the a restriction graph $\m{W}$). In this work we do not consider how to extend the construction of the structure matrix to include a restriction graph, although we assume this can be done (and would be an interesting matter for further research). Still we can define the following graphs that indicate the invariant and non-invariant positions in any given edge-type {class} $\rvec{T}(\vec{r}, \vec{c}, \m{W})$: }
{
\begin{dfn} \label{dfn:Wnon-invariant_invaraint}
For a given edge-type class $\rvec{T}(\vec{r}, \vec{c}, \m{W})$ ($T = (\vec{r}, \vec{c}, \m{W})$) we define three directed graphs over $[n]$:
\begin{itemize}
\item $\m{W}_{T, non-invariant}$  contains edges $i \gto{} j$ if and only if they are non-invariant edges of the edge-type class.
\item $\m{W}_{T, invariant-1-position}$ contains edges $i \gto{} j$ if and only if they are 1-position invariant edges of the edge-type class, meaning edges that exists in every graph of the edge-type class {$\rvec{T}(\vec{r}, \vec{c}, \m{W})$}.
\item $\m{W}_{T, invariant-0-position}$ contains edges $i \gto{} j$ if and only if they are 0-position invariant edges of the edge-type class, meaning edges that do not exists in any of the graphs of the edge-type class {$\rvec{T}(\vec{r}, \vec{c}, \m{W})$}.
\end{itemize}
\end{dfn}
As mentioned, we do not know how to precisely construct the above defined graphs, however that is immaterial for our purposes.
}

More recently, Barvinok \cite{barvinok2010} (and later Barvinok and Hartigan~\cite{barvinok2013number} using the graph vernacular, see also Chatterjee \etal~\cite{chatterjee2011random}) considered edge-types from a somewhat different angle that is more suitable for our purposes; in fact, {Barvinok's} results form the cornerstone of this work. We go through some of his main results.

\begin{thm}[Theorem 5 \cite{barvinok2010}] \label{thm:BarvinokThm1}
Let $T=(\vec{r},\vec{c}, {\m{W}})$ be an edge-type of some graph. Define the function $Q_T:\mathbb{R}^n\times \mathbb{R}^n \to \mathbb{R}$
\begin{align}\label{eq:gen_func}
Q_T(\vec{x}, \vec{y}) = \left( \prod_{i=1}^{n} x_i^{-\vec{r}(i)} \right) \left( \prod_{j=1}^{n} y_j^{-\vec{c}(j)} \right) \prod_{i,j} (1 + {\left\{\mat{A}_{\m{W}}\right\}_{ij}} x_i y_j ),
\end{align}
for $\rvecr{x} = (x_1, x_2, \ldots, x_n)$ and $\rvecr{y} = (y_1, y_2, \ldots, y_n)$,
and let
\begin{align}
\alpha(T) = \inf_{\vec{x},\vec{y}\in \mathbb{R}^n_+} Q_T( \vec{x}, \vec{y}).\label{eq:opt1}
\end{align}
Then given a directed graph $\m{G}$ {and a restriction graph $\m{W}$}, for the cardinality of the corresponding edge-type class, $| \rvec{T}(\m{G}, {\m{W}}) |$, we have that
\begin{align} \label{eq:thm1Barvinok}
2^{-\gamma n\log{n}}\alpha(T_{\m{G},{\m{W}}}) \leq |\rvec{T}(\m{G},{\m{W}})| \leq \alpha(T_{\m{G},{\m{W}}})
\end{align}
where $\gamma > 0$ is a universal constant.
\end{thm}
This bound is typically tight in the exponential order, as many edge-type classes are of size $2^{\Omega(n^2)}$\footnote{$f(n) = \Omega( g(n))$ according to the Landau notation in complexity denotes the case when $f(n)$ is bounded from below by $g(n)$ asymptotically, meaning when $\liminf_{n \to \infty} \frac{f(n)}{g(n)} > 0$}. Indeed, note that the total number of (labeled, directed) graphs on $n$ vertices is $2^{n^2}$, while the number of edge-types cannot exceed $2^{2n\log{n}}$ (recall that either $\vec{r}$ or $\vec{c}$ may take one of $(n+1)^n$ values, thus, we have $(n+1)^{2n}$ which leads to this bound, since not all pairs $(\vec{r}, \vec{c})$ are valid ones).

Barvinok observed \cite{barvinok2010} that by a simple substitution $x_i = e^{s_i}$, $y_i = e^{t_i}$ in $Q_T( \rvecr{x}, \rvecr{y})$ one obtains the following function
\begin{align} \label{eq:BarvinokDualG}
G_T( \rvecr{s}, \rvecr{t}) & = \log \left( Q_T( \rvecr{x}, \rvecr{y}) \right) \nonumber \\
& = - \sum_{i=1}^{n} \vec{r}(i) s_i - \sum_{j=1}^{n} \vec{c}(j) t_j + \sum_{i,j} \ln (1 + {\left\{\mat{A}_{\m{W}}\right\}_{ij}} e^{s_i + t_j})
\end{align}
for $\rvecr{s} = (s_1, s_2, \ldots, s_n)$ and $\rvecr{t} = (t_1, t_2, \ldots, t_n)$.
As the above optimization problem is convex on $\mathcal{R}^{n^2}$, the computation of the infimum can be done using efficient optimization algorithms.

Now, we require the following two definitions:
\begin{dfn}[Entropy of a Random Graph \cite{barvinok2010}] \label{dfn:entropyOfMatrix}
Given a random graph $\m{F}\sim \{p_{ij}\}$ the entropy is given as follows:
\begin{align}
H( \m{F} ) = \sum_{i,j} H_b(p_{ij})
\end{align}
where $H_b(\cdot)$ denotes the binary entropy function.
\end{dfn}

\begin{dfn}[Maximum Entropy Random Graph \cite{barvinok2010}] \label{dfn:maximumEntropyGraph}
We associate with every edge-type $T = {(\vec{r}, \vec{c}, \m{W})}$ a {\em maximum entropy random graph} $\m{F}_T$, which is the unique random graph that maximizes the entropy out of all {$\m{W}$ constrained} random graphs (with independent edges) whose expected edge-type is $T$. Explicitly:
\begin{align}\label{eq:max_ent_graph}
  & \m{F}_T \triangleq \argmax_{\m{F}\sim\{p_{ij}\}} H(\m{F}) \nonumber \\
  & \text{s.t.} \quad \Expt T_{\m{F}} = T { \textrm{ and } p_{ij} = 0 \textrm{ whenever } \left\{ \mat{A}_{\m{W}} \right\}_{i,j} = 0}.
\end{align}
Note that the constraints $\Expt  T_{\m{F}} = T (= (\vec{r}, \vec{c}, {\m{W}} ))$ and {$p_{ij} = 0$ whenever $\left\{ \mat{A}_{\m{W}} \right\}_{i,j} = 0$} define a polytop over random graphs
$\m{F}\sim \{p_{ij}\}$ such that
\begin{align}
& \sum_{j=1}^{n} p_{ij} = \vec{r}(i), \quad \forall i = 1, 2, \ldots, n \nonumber \\
& \sum_{i=1}^{n} p_{ij} = \vec{c}(j), \quad \forall j = 1, 2, \ldots, n { \textrm{ and }} \nonumber \\
& {p_{ij} = 0 \textrm{ whenever } \left\{ \mat{A}_{\m{W}} \right\}_{i,j} = 0}.
\end{align}
We denote this poyltop either by $\mathcal{P}( \vec{r}, \vec{c}, {\m{W}})$ or $\mathcal{P}( T )$. The \emph{maximum entropy random graph} can be alternatively defined as maximizing the entropy over this polytop.
\end{dfn}

The uniqueness of $\m{F}_T$ is guaranteed from the fact that $H( \m{F})$ is strictly concave. Finally, note that Barvinok \cite{barvinok2010} referred to $\m{F}_T$ as the \emph{maximum entropy matrix} and denoted it as $\mat{Z}$.

{\begin{rem}
Note that for all $\m{F} \sim \{ p_{ij} \} \in \mathcal{P}( \vec{r}, \vec{c}, {\m{W}})$
\begin{align}
\textrm{ if } \left\{ \mat{A}_{\m{W}_{T,invariant-1-position}} \right\}_{i,j} = 1 \quad \textrm{{then} } p_{ij} = 1
\end{align}
and
\begin{align}
\textrm{ if } \left\{ \mat{A}_{\m{W}_{T,invariant-0-position}} \right\}_{i,j} = 1 \quad \textrm{{then} } p_{ij} = 0
\end{align}
with $T = (\vec{r}, \vec{c}, {\m{W}} )$. This follows from the explanation of the assumption of a non-empty polytop in \cite[Lemma 2]{barvinok2010}. A non-empty polytop requires that there exists a random graph such that $p_{ij} \in (0,1)$ for all $i \in [n]$ and $j \in [n]$. The explanation provided by Barvinok \cite[bottome of page 319]{barvinok2010} is that the random graph can be viewed as an average over a subset of $\m{G} \in \rvec{T}( \vec{r}, \vec{c}, \m{W})$. Given this interpretation, invariant positions result with probabilities $1$ (in the case of invariant 1-position) and $0$ (in the case of invariant 0-position).
\end{rem}
}

Given these definitions Barvinok \cite{barvinok2010} provided a more instructive expression for the cardinality of the edge-type class that admits a natural interpretation in the language of compression.
\begin{lem}[Lemma 6 \cite{barvinok2010}]\label{lem:BarvinokDual}
Suppose that the polytop $\mathcal{P}( T )$, for some edge-type $T = (\vec{r}, \vec{c}, {\m{W}})$, has a non-empty interior, that is, contains a random graph $\m{F} \sim {\{p_{ij}\}}$ such that $0 < p_{ij} < 1$ for all $i,j$ {whenever $\mat{A}_{\m{W}_{i,j}} = 1$ and zero otherwise}. For {a} maximum entropy random graph $\m{F}_T \sim {\{p_{ij}\}}$ we have
\begin{align} \label{eq:lem:BarvinokDualEquationZ}
p_{ij} = \frac{ \xi_i \eta_j}{1 + \xi_i \eta_j}, \quad \forall i,j
\end{align}
for some vectors {$\rvecr{x}^\star = (\xi_1, \xi_2, \ldots, \xi_{n})$ and $\rvecr{y}^\star = (\eta_1, \eta_2, \ldots, \eta_{n} )$.}
Moreover,
\begin{align} \label{eq:lem:BarvinokDualEquationZ2}
\alpha(T) = e^{H(\m{F}_T)}.
\end{align}
Conversely, if the infimum $\alpha( T )$ in Theorem \ref{thm:BarvinokThm1} is attained at a certain point $\rvecr{x}^\star = (\xi_1, \xi_2, \ldots, \xi_{n})$ and $\rvecr{y}^\star = (\eta_1, \eta_2, \ldots, \eta_{n} )$ then for the maximum entropy random graph $\m{F}_T \sim {\{p_{i,j}\}}$ equations (\ref{eq:lem:BarvinokDualEquationZ}) and (\ref{eq:lem:BarvinokDualEquationZ2}) hold.
\end{lem}
Namely, from Theorem \ref{thm:BarvinokThm1} and Lemma \ref{lem:BarvinokDual} (equations \eqref{eq:thm1Barvinok} and \eqref{eq:BarvinokDualG}) the number of nats (bits) required to represent an edge-type class is roughly the maximum entropy random graph associated with the edge-type. Moreover, the distribution of the maximum entropy random graph takes a specific special form, as shown in \eqref{eq:lem:BarvinokDualEquationZ}. 

{
\begin{rem}
The assumption of a non empty interior of the polytop $\mathcal{P}(T)$ is explained by Barvinok in \cite{barvinok2010} as follows: \emph{The condition that the polytop $\mathcal{P}(\vec{r}, \vec{c}, \m{W})$ has a non-empty interior is equivalent to the requirement that for every choice of $k \in [n]$ and $\ell \in [n]$ there is a graph $\m{G} \in \rvec{T}(\vec{r}, \vec{c}, \m{W})$, such that $k \ngto{G} \ell$ and there is a graph $\m{H} \in \rvec{T}(\vec{r},\vec{c}, \m{W})$ such that $k \gto{H} \ell$.
One can take $\m{F}$ to be the average of all graphs $\m{D} \in \rvec{T}(\vec{r},\vec{c}, \m{W})$. In other words, we require the
set $\rvec{T}(\vec{r},\vec{c}, \m{W})$ to be reasonably large.} (where we made the required adjustments to refer to graphs instead of matrices).\newline
We feel that this is indeed not a sufficient consideration of the matter, specifically due to the existence of the additional restriction graph $\m{W}$. In other words, the proof already excludes specific edges. The same mechanism can be applied to invariant edges. Definition \ref{dfn:Wnon-invariant_invaraint} provides additional restriction {matrices} {graphs} that mark the invariant positions. Invariant 0-positions are exactly like the restricted positions marked by $\m{W}$ as non-existing, since they define non-existing edges. Invariant 1-positions require us to adjust the vectors $\vec{r}$ and $\vec{c}$ appropriately once they are excluded. Given this, the proof of the above can be extended as follows, assuming the condition in Lemma \ref{lem:requirementOnRestrictionGraph} holds:
\begin{itemize}
\item Create an extended restriction graph $\hat{\m{W}} = \m{W} \wedge \bar{\m{W}}_{T, invariant-1-position} \wedge \bar{\m{W}}_{T, invariant-0-position}$
\item For every edge $i \gto{} j$ in $\m{W}_{T, invariant-1-position}$ update $\vec{r}(i) = \vec{r}(i) -1 $  and $\vec{c}(j) = \vec{c}(j) - 1$
\item Consider $\rvec{T}( \vec{r}, \vec{c}, \hat{\m{W}})$ in the proof of {Theorem} {Lemma} \ref{lem:BarvinokDual}
\end{itemize}
\end{rem}
}

The maximum entropy 
interpretations lead to an additional observation by Barvinok \cite{barvinok2010}, in the spirit of the method-of-types: The maximum entropy random graph $\m{F}_T$ is uniformly distributed over the corresponding edge-type class $\rvec{T}$. This result is given next:
\begin{thm}[Theorem 8 \cite{barvinok2010}] \label{thm:Barvinok4}
Let $T = (\vec{r}, \vec{c}, \m{W})$ be any edge-type over $[n]$ such that the polytop $\mathcal{P}( T )$ has a non-empty interior and let $\m{F}_T \in \mathcal{P}(T)$ be the maximum entropy random graph.
Then the probability mass function of $\m{F}_T$ is constant on the set $\rvec{T}( \vec{r}, \vec{c}, \m{W})$, and moreover,
\begin{align}
\Pr \left\{ \m{F}_T = \m{D} \right\} = e^{- H (\m{F}_T )}, \quad \forall \m{D} \in \rvec{T}(\vec{r}, \vec{c}, \m{W}).
\end{align}
\end{thm}
Given the above result, with some abuse of notation, we will denote $\rvec{T}(\vec{r}, \vec{c}, \m{W})$, that is, an $(\vec{r}, \vec{c})$-type restricted by the graph $\m{W}$, also using its maximum entropy random graph $\m{F}_T$ as $\rvec{T}(\m{F}_T, {\m{W}})$.

\section{Method of Type Extended} \label{sec:MethodOfTypesExtended}
The first contribution of this work is the extension of Barvinok's results \cite{barvinok2010} in the spirit of the {method of types}.
Barvinok provides two important observations:
\begin{enumerate}

\item The cardinality of an edge-type is approximately exponential in the entropy of the maximum entropy random graph.
\item Given a random graph distributed according to the maximum entropy random graph distribution its probability across the corresponding edge-type is uniform and is thus $e^{- H (\m{F}_T )}$.

\end{enumerate}

Although these results are very much in the flavor of the method of types, they lack several aspects and must be considerably extended in order for us to fully follow through and obtain results similar to those obtained using the method of types. First, when considering the probability of obtaining a graph from a specific edge-type, they consider only the maximum entropy random graph distribution. Second, similar to the method of types, edge-types need to be perturbed slightly in order to obtain the high probability results required.

In this section we consider both extensions.

\subsection{A Family of Random Graphs} \label{ssec:FamilyOfRandomGraphs}
As said, the observation of Barvinok \cite{barvinok2010} regarding the probability of obtaining a graph from a specific edge-type is limited to random graphs distributed according to the corresponding maximum entropy random graph distribution. We extend upon this. {We begin by defining the following set of distributions:}
\begin{dfn} \label{dfn:familyDist}
Let $\mathbb{D}(n, \m{W})$ denote a family of random graphs, $\m{F} \sim {\{ p_{ij} \}}$, over $[n]$, restricted by the graph $\m{W}$, for which the probabilities are:
\begin{align}
p_{ij} = 0, \quad \textrm{if } \left\{\mat{A}_{\m{W}} \right\}_{i,j} = 0
\end{align}
and
\begin{align}
p_{ij} = \frac{e^{-a_i}e^{-b_j}}{1 + e^{-a_i}e^{-b_j}}
\end{align}
otherwise, for some $a_1, a_2, \ldots, a_n$ and $b_1, b_2, \ldots, b_n$.
\end{dfn}

For this family of random graphs we have the following result:
\begin{thm} \label{thm:probabilityOfrandomGraphInEdgeType}
For any $\m{F} \sim {\{ p_{ij} \}} \in \mathbb{D}(n, \m{W})$ and any edge-type $T = (\vec{r}, \vec{c}, \m{W})${, if $\m{F}_T \ll \m{F}$}, for all $\m{G} \in \rvec{T}( \vec{r}, \vec{c}, \m{W})$ we have that
\begin{align}
\Pr \left( \m{F} =  \m{G} \right) = e^{-H (\m{F}_T) - \sum_{i,j:\left\{ \mat{A}_{\m{W}} \right\}_{i,j} = 1} D( (p_T)_{i,j} || p_{ij} ) },
\end{align}
where $\m{F}_T \sim { \{ (p_T)_{i,j} \}}$ is the maximum entropy random graph over the polytop $\mathcal{P}(\vec{r}, \vec{c}, \m{W})$.
\end{thm}
\begin{IEEEproof}
The proof is given in Appendix \ref{appendix:thm:probabilityOfrandomGraphInEdgeType}.
\end{IEEEproof}

The importance of the above result is twofold. First, as we can clearly see, for \emph{any} distribution in $\mathbb{D}(n, \m{W})$ we have that the probability is uniform over any given edge-type $\rvec{T}( \vec{r}, \vec{c}, \m{W})$ and second it depends on the KL-divergence between $\m{F}$ and the maximum entropy random graph corresponding to that edge-type. This is very similar to the standard results from the method-of-types, giving an additional important interpretation for $\m{F}_T$.

{
\begin{rem}
The condition $\m{F}_T \ll \m{F}$ does not come to suggest that if the condition is not {held} the probability $\Pr \left( \m{F} = \m{G} \right)$ is zero. This is not the case. However, if the condition is not held, stating the probability as a function of $\m{F}_T$ is problematic. This can be circumvented. The condition is relevant to pairs $(i,j)$ for which $p_{ij} = 1$ (or $p_{ij} = 0$) ($\m{F} \sim \{ p_{ij} \}$). If this is the case, we may have a non-zero probability {for $\Pr \left( \m{F} = \m{G} \right)$} only if $i \gto{G} j$ (or $i \ngto{G} j$, respectively). We can thus consider an extended restriction graph $\m{W}$ that will exclude such edges. If these are existing edges their contribution can be reduced from the $(\vec{r}, \vec{c})$ pair. By doing this we consider a different edge-type, and a random graph in which the probabilities are {all} in the region of $(0,1)$.
\end{rem}
}
{The set $\mathbb{D}(n, \m{W})$ is a set of discrete distribution points. It is a subset of the general set of random graphs $\m{F} \sim \{ p_{ij} \}$, meaning random directed graphs in which each edge exists independently of any other edge, with a predetermined probability. However, convex combinations over the set $\mathbb{D}(n, \m{W})$ can provide any random graph from the general set, as shown in the next claim:}
{
\begin{lem} \label{lem:extengingTheSetD}
For any random graph $\m{F} \sim \{ p_{ij} \}$ there exists a $K$ and a vector $\vec{\lambda}$ of length $K$ with $\sum_{k=1}^K \vec{\lambda}(k) = 1$, and  $\vec{\lambda}(k) \in (0, 1]$ for all $k \in [K]$, and a set of $K$ random graphs in $\mathbb{D}(n, \m{W})$ denoted as $\{ \m{F}_1, \m{F}_2, \ldots, \m{F}_K \}$, $\m{F}_k \sim \{ p_{ij}^k\}$ for all $k \in [K]$ such that
\begin{align} \label{eq:lem:extendingTheSetD}
\m{F} \sim \left\{ \sum_{k =1}^K \vec{\lambda}(k) p_{ij}^k \right\}.
\end{align}
The other direction also holds, meaning that taking any combination like the one given above will result with a random graph from the general set.
\end{lem}
\begin{IEEEproof}
The proof is given in Appendix \ref{appendix:lem:extendingTheSetD}.
\end{IEEEproof}}


{From this observation we may extend the result of Theorem \ref{thm:probabilityOfrandomGraphInEdgeType} as follows:
\begin{thm} \label{thm:boundsOnProbabilityGeneralDistribution}
For any random graph $\m{F} \sim \{ p_{ij} \}$ and any edge-type $T = (\vec{r}, \vec{c}, \m{W})$, we have that
\begin{align}
e^{-H( \m{F}_T) - \sum_{k=1}^K \vec{\lambda}(k) \sum_{i,j:\left\{ \mat{A}_{\m{W}} \right\}_{i,j} = 1} D( (p_T)_{i,j} || p_{ij}^k )} \leq \Pr \left( \m{F} = \m{G} \right)
\end{align}
if there exists a value $K$, a set of random graphs $\{\m{F}_1, \ldots, \m{F}_K \}$, $\m{F}_k  \sim \{ p_{ij}^k\} \in \mathbb{D}(n , \m{W})$, $\m{F}_T \ll \m{F}_k$ for all $k \in [K]$, and the vector $\vec{\lambda}$ such that
\begin{align}
\m{F} \sim \left\{ \sum_{k =1}^K \vec{\lambda}(k) p_{ij}^k \right\}.
\end{align}
\end{thm}
\begin{IEEEproof}
The proof is given in Appendix \ref{appendix:thm:boundsOnProbabilityGeneralDistribution}.
\end{IEEEproof}
}

From {Theorem \ref{thm:probabilityOfrandomGraphInEdgeType}} we may also consider the probability of picking any graph within a specific edge-type class:
\begin{thm} \label{thm:probabilityOfAnEdgeType}
For any $\m{F} \sim { \{p_{ij}\}} \in \mathbb{D}(n, \m{W})$ and any $\rvec{T}(\vec{r}, \vec{c}, \m{W})$, with maximum entropy random graph $\m{F}_T$, if $\m{F}_T \ll \m{F}$,
\begin{align}
n^{-2\gamma(2n)} e^{- \sum_{i,j:\left\{\mat{A}_{\m{W}} \right\}_{i,j}=1} D( (p_T)_{i,j} || p_{ij} ) } \leq \Pr \left( \m{F} \in \rvec{T}( \vec{r}, \vec{c}, \m{W}) \right) \leq e^{- \sum_{i,j:\left\{\mat{A}_{\m{W}} \right\}_{i,j}=1} D( (p_T)_{i,j} || p_{ij} ) }.
\end{align}
for some absolute constant $\gamma > 0$.
\end{thm}
\begin{IEEEproof}
\begin{align}
\Pr \left( \m{F} \in \rvec{T}( \vec{r}, \vec{c}, \m{W}) \right)
& = \sum_{ \m{G} \in \rvec{T}( \vec{r}, \vec{c}, \m{W}) } \Pr \left( \m{F} = \m{G} \right) \nonumber \\
& = \sum_{ \m{G} \in \rvec{T}( \vec{r}, \vec{c}, \m{W}) } e^{-H (\m{F}_T) - \sum_{i,j:\left\{\mat{A}_{\m{W}} \right\}_{i,j}=1} D( (p_T)_{i,j} || p_{ij} ) }  \nonumber \\
& = \left| \rvec{T}( \vec{r}, \vec{c}, \m{W}) \right| e^{-H (\m{F}_T) - \sum_{i,j:\left\{\mat{A}_{\m{W}} \right\}_{i,j}=1} D( (p_T)_{i,j} || p_{ij} ) }.
\end{align}
Using the results of Theorem \ref{thm:BarvinokThm1} and Lemma \ref{lem:BarvinokDual} we obtain the desired result.
\end{IEEEproof}
From the above results it is evident that for any $\m{F} \sim { \{p_{ij} \}} \in \mathbb{D}(n, \m{W})$ such that $\m{F} \neq \m{F}_T$ we have that the probability of $\rvec{T}( \vec{r}, \vec{c}, \m{W})$ under $\m{F}$ is exponentially small (for large enough $n$). When $\m{F} = \m{F}_T$ the above results are not very informative. 

Another immediate result is an extension of Sanov's result.
\begin{thm} \label{thm:ExtensionSanov}
Consider a set
\begin{align} \label{eq:thm:extensionSanov_defSet}
A = \{ \m{G} : \m{G} \in \cup_{\m{F}_T \in \bold{F}} \rvec{T}( \m{F}_T, {\m{W}} )  \}
\end{align}
where $\bold{F}$ is a {subset of the} set of maximum entropy random graphs. {$\m{F}_T$ denotes a member of this subset, meaning some maximum entropy random graph.}
Assume a random graph $\m{F} \sim { \{ p_{ij} \}} \in \mathbb{D}(n, \m{W})$. The probability of the set $A$ with respect to $\m{F}$ can be bounded as follows:
\begin{align} \label{eq:thm:extensionSanov}
e^{-\gamma 4 n \log(n) -\min_{\m{F}_T \in \bold{F} } \sum_{i,j:{\{ \mat{A}_{\m{W}} \}_{i,j}=1}} D( {(p_T)_{i,j}} || {p_{ij}} ) } \leq {\Pr}_{\m{F}} \left( A \right) \leq  e^{2 n \log(n+1) -\min_{\m{F}_T \in \bold{F} } \sum_{i,j:{\{ \mat{A}_{\m{W}} \}_{i,j}=1}} D( {(p_T)_{i,j}} || {p_{i,j}} ) }
\end{align}
{where $\gamma >0$ is some universal constant.}
For large enough $n$, {if the following condition holds:}
\begin{align} \label{eq:SanovConditionOnKLdivergence}
\min_{\m{F}_T \in \bold{F} } \sum_{i,j:{\{ \mat{A}_{\m{W}} \}_{i,j}=1}} D( {(p_T)_{i,j}} || {p_{ij}} ) = \Theta( n^2)
\end{align}
where $\Theta$ is the Landau notation\footnote{$f(n) = \Theta(n^2)$ means that there exist $k_1, k_2 >0$ and $n_0$ such that $k_1 n^2 \leq f(n) \leq k_2 n^2$ for all $n > n_0$.},
the above probability can be approximated by
\begin{align}
\Pr ( A ) \approx  e^{ -\min_{\m{F}_T \in \bold{F} } \sum_{i,j:{\{ \mat{A}_{\m{W}} \}_{i,j}=1}} D( {(p_T)_{i,j}} || {p_{ij}} )}
\end{align}
where the approximation is in the normalized logarithmic sense\footnote{The approximation is in the normalized logarithmic sense, meaning that $A \approx B$ if $\lim_{n \to \infty} \frac{1}{n^2} \log \frac{A}{B} = 0$.}.
\end{thm}
\begin{IEEEproof}
The proof is given in Appendix \ref{appendix:thm:ExtensionSanov}.
\end{IEEEproof}

\begin{rem}{Note that not any random graph is a viable maximum entropy random graph. This can be seen from the fact that the number of possible random graphs is infinite, while the number of edge-{type} classes for a given $[n]$ is not.}
\end{rem}

\subsection{Delta Edge-Types} \label{ssec:DeltaEdgeType}
As noted above, when $\m{F} = \m{F}_T$ we still do not have an informative claim regarding the probability of the corresponding edge-type. In other words, {we cannot guarantee with probability one that a realization of a maximum entropy random graph will be in the corresponding edge type class.} In order to resolve this we extend the definition of an edge-type similarly to the extension done in the \emph{method of type} sometimes referred to as \emph{strong typicality}. {Before going into this extension we require the following definition:
\begin{dfn} \label{dfn:density}
A directed graph $\m{G}$ over $[n]$ is said to be of degree-density $\bold{d}_{\m{G}}(n)$ if and only if all of its incoming degrees and all of its outgoing degrees are of order $\bold{d}_{\m{G}}(n)$ in the following sense:
\begin{align}
\vec{r}(i) & = \Theta( \bold{d}_{\m{G}}(n)) \quad \forall i \in [n] \nonumber \\
\vec{c}(j) & = \Theta( \bold{d}_{\m{G}}(n)) \quad \forall j \in [n].
\end{align}
The above definition allows us to distinguish between the sparsity level of graphs, as long as it holds for each of the incoming and outgoing degrees. Moreover, note that the degree-density property is a property of the edge-type class, as such we also use the notation $\bold{d}_{T}(n)$.
\end{dfn}
Given the above definition we can consider the following:}
\begin{dfn} \label{dfn:strongTypicalityEdgeType}
We say that the graph $\m{G}$ over $[n]$ has $\delta$ edge-type $T = (\rvecr{r}, \rvecr{c}, \m{W})$ if {$\m{G} \wedge \m{W} = \m{G}$} and
\begin{align}
\frac{1}{{\bold{d}_{T}(n)}} \left| \rvecr{r}_{\m{G}}(i) - \rvecr{r}(i) \right| & < \delta \quad \forall i \in [n] \nonumber \\
\frac{1}{{\bold{d}_{T}(n)}} \left| \rvecr{c}_{\m{G}}(i) - \rvecr{c}(i) \right| & < \delta \quad \forall i \in [n]
\end{align}
{where $\vec{r}_{\m{G}}$ and $\vec{c}_{\m{G}}$ are the outgoing and incoming degrees of $\m{G}$ respectively. }
The set of such graphs is denoted as $\rvec{T}_{\delta}(\rvecr{r}, \rvecr{c}, \m{W})$, or using the maximum entropy random graph {over $\mathcal{P}(\vec{r}, \vec{c}, \m{W})$}, as $\rvec{T}_{\delta}( \m{F}_T, {\m{W}})$.
\end{dfn}

\begin{rem} \label{rem:delta_edge_disjoint_union}
Note that similar to $P$-typical sets in \cite[Definition 2.8]{csiszar_korner} the above defined set is a union of sets $\rvec{T}(\tilde{\rvecr{r}}, \tilde{\rvecr{c}},\m{W})$ for {all pairs} ($\tilde{\rvecr{r}}$, $\tilde{\rvecr{c}}$) that satisfy $\frac{1}{{\bold{d}_{T}(n)}}| \tilde{\rvecr{r}}(i) - \rvecr{r}(i) | < \delta$ for all $i \in [n]$ and $\frac{1}{{\bold{d}_{T}(n)}}| \tilde{\rvecr{c}}(j) - \rvecr{c}(j) | < \delta$ for all $j \in [n]$. {This is a disjoint union, since each such pair ($\tilde{\rvecr{r}}$, $\tilde{\rvecr{c}}$) defines a unique edge-type class $\rvec{T}(\tilde{\rvecr{r}}, \tilde{\rvecr{c}},\m{W})$.}  {We sometimes say that $\tilde{\rvecr{r}}$ is $\delta$-close to $\vec{r}$, and $\tilde{\rvecr{c}}$ is $\delta$-close to $\vec{c}$.}
\end{rem}

Given this definition of $\delta$ edge-types we obtain the following result:
\begin{lem} \label{lem:extensionLemma2.12_CK}
For any $n$, every $\rvec{T}_{\delta}(\rvecr{r}, \rvecr{c}, \m{W})$ for some small $\delta > 0$ we have that
\begin{align}
\Pr \left( \rvec{T}_{\delta}( \rvecr{r}, \rvecr{c}, \m{W}) \right) \geq 1 - 4 { n e^{-2  \frac{\bold{d}_{T}(n)^2 \delta^2}{n} } }   \nonumber
\end{align}
where the probability {can be according to any random graph from the polytop $\mathcal{P}(\vec{r},\vec{c}, \m{W})$, and $T = (\vec{r},\vec{c}, \m{W})$}.
\end{lem}
\begin{IEEEproof}
Consider the following sets of ``bad'' events:
\begin{align}
E_i^{\rvecr{r}} & = \left\{ \frac{1}{{\bold{d}_{T}(n)}} \left| \rvecr{r}_{\m{G}}(i) - \rvecr{r}(i) \right| \geq \delta \right\}, \quad \forall i \in [n] \nonumber\\
E_{{j}}^{\rvecr{c}} & = \left\{ \frac{1}{{\bold{d}_{T}(n)}} \left| \rvecr{c}_{\m{G}}({j}) - \rvecr{c}({j}) \right| \geq \delta \right\}, \quad \forall {j} \in [n].
\end{align}
{
We may rewrite these events as follows:
\begin{align}
E_i^{\rvecr{r}} & = \left\{ \frac{1}{n} \left| \rvecr{r}_{\m{G}}(i) - \rvecr{r}(i) \right| \geq  \frac{\bold{d}_{T}(n) \delta}{n} \right\}, \quad \forall i \in [n] \nonumber\\
E_{{j}}^{\rvecr{c}} & = \left\{ \frac{1}{n} \left| \rvecr{c}_{\m{G}}({j}) - \rvecr{c}({j}) \right| \geq \frac{\bold{d}_{T}(n) \delta}{n} \right\}, \quad \forall {j} \in [n].
\end{align}
We do so because each $\vec{r}_{\m{G}}(i)$ and each $\vec{c}_{\m{G}}(j)$ for all $i \in [n]$ and $j \in [n]$ are each a summation of $n$ independent random variables (binary random variables - existence or non-existence of an edge). To apply Hoeffding's inequality we compare their empirical mean to the expectation, thus we must divide by the number of random variables - $n$.}
For each such event we {can now} apply Hoeffding's inequality to obtain the following bound on the probability:
\begin{align}
\Pr \left( E_i^{\rvecr{r}} \right) & \leq 2 e^{-2 n \left( \frac{\bold{d}_{T}(n) \delta}{n} \right)^2} = 2 e^{-2  \frac{\bold{d}_{T}(n)^2 \delta^2}{n} } \nonumber \\
\Pr \left( E_j^{\rvecr{c}} \right) & \leq 2 e^{-2 n \left( \frac{\bold{d}_{T}(n) \delta}{n} \right)^2} = 2 e^{-2  \frac{\bold{d}_{T}(n)^2 \delta^2}{n} }
\end{align}
for all $i \in [n]$ {and all $j \in [n]$}. Using the union bound on the probability that at least one of the events happens:
\begin{align}
\Pr \left( \cup_{i} E_i^{\rvecr{r}} \cup \cup_j E_j^{\rvecr{c}} \right) & \leq \sum_{i} \Pr\left( E_i^{\rvecr{r}} \right) + \sum_j \Pr \left( E_j^{\rvecr{c}} \right) \nonumber \\
& \leq {2 n e^{-2  \frac{\bold{d}_{T}(n)^2 \delta^2}{n} } + 2 n e^{-2  \frac{\bold{d}_{T}(n)^2 \delta^2}{n} }}.
\end{align}
The above bound holds for any $n$ and any $\delta > 0$. This concludes the proof.
\end{IEEEproof}
{From the above result we can see that when $\bold{d}_T(n) = \Theta(n)$, for large enough $n$ (and $\delta > 0$) such that $n \delta^2$ is large, we have that a realization of a random graph from $\mathcal{P}(\vec{r}, \vec{c}, \m{W})$ will obtain a graph in $\rvec{T}_{\delta}( \rvecr{r}, \rvecr{c}, \m{W})$, with probability close to one. However, our bound does not reach the same conclusion when the degree-density is reduced. }

Next, we consider the cardinality of a $\delta$ edge-type class, $\rvec{T}_{\delta}( \m{F}_T, {\m{W}})$, for $T = \left( \rvecr{r}, \rvecr{c}, \m{W} \right)$.
\begin{lem} \label{lem:strongTypicalityCardinality}
For a given $T = \left( \rvecr{r}, \rvecr{c}, \m{W} \right)$, the cardinality of $\rvec{T}_{\delta}( \m{F}_T, {\m{W}} )$ with constant $\delta \in \left( 0, \frac{1}{2} \right]$, and the universal constant $\gamma > 0$ defined in Theorem \ref{thm:BarvinokThm1}, has the following property:
\begin{align}
- \frac{\gamma \log n}{n} \leq \frac{1}{n^2}\log \left| \rvec{T}_{\delta}( \m{F}_T, {\m{W}} ) \right| - \frac{1}{n^2} H( \m{F}_T ) \leq H_b(\delta) + \frac{\log n {\bold{d}_T(n)}}{n^2}.
\end{align}
Thus, for sufficiently large $n$ we have that
\begin{align}
\frac{1}{n^2}  \left| \log \left| \rvec{T}_{\delta}( \m{F}_T, {\m{W}} ) \right| - H( \m{F}_T ) \right| \leq H_b( \delta) + \epsilon_n
\end{align}
where $\epsilon_n \to 0$ as $n \to \infty$.
\end{lem}
\begin{IEEEproof}
A graph $\m{G} \in \rvec{T}_{\delta}( \m{F}_T )$ has an outgoing degree that differs from $\rvecr{r}(i)$ by at most ${\bold{d}_T(n)} \delta$ for every $i \in [n]$, from the set of $n$ possible outgoing edges. Extending this logic to the entire set of outgoing edges, $n^2$ possible edges, meaning the $n^2$ {possible} outgoing edges may differ by at most {$n \bold{d}_T(n) \delta$} from the edges of a source graph. Given this observation we have that the cardinality $\left| \rvec{T}_{\delta}( \m{F}_T ) \right|$ can be lower and upper bounded as follows:
\begin{align}
\left| \rvec{T}(\m{F}_T, {\m{W}}) \right| \leq \left| \rvec{T}_{\delta}( \m{F}_T, {\m{W}} ) \right| & \leq \left| \rvec{T}(\m{F}_T, {\m{W}}) \right|  \sum_{k=0}^{n {\bold{d}_T(n)} \delta} \binom{n^2}{k}
\end{align}
since $\rvec{T}_{\delta}{( \m{F}_T, \m{W} )}$ contains at least $\rvec{T}{( \m{F}_T, \m{W} )}$ but may contain every possible change of at most $n {\bold{d}_T(n)} \delta$ outgoing edges, if those create a valid graph (meaning that both the outgoing and the ingoing degrees actually comply with the constraints on every $i \in [n]$).
Note that similarly we could have chosen to examine the degrees of the incoming edges. Using Theorem \ref{thm:BarvinokThm1} and Lemma \ref{lem:BarvinokDual} we have that
\begin{align}
2^{-\gamma n \log n} e^{H( \m{F}_T )} \leq \left| \rvec{T}_{\delta}( \m{F}_T, {\m{W}} ) \right| & \leq e^{H( \m{F}_T )} \sum_{k=0}^{n {\bold{d}_T(n)} \delta} \binom{n^2}{ k} \nonumber \\
2^{-\gamma n \log n} \leq \frac{\left| \rvec{T}_{\delta}( \m{F}_T, {\m{W}} ) \right|}{e^{H( \m{F}_T )} } & \leq \sum_{k=0}^{n {\bold{d}_T(n)} \delta} \binom{n^2}{ k} \stackrel{a}{\leq} \sum_{k=0}^{n {\bold{d}_T(n)} \delta} e^{n^2 H_b(\alpha_k)} \stackrel{b}{\leq} n {\bold{d}_T(n)} e^{n^2 H_b(\delta)} = e^{n^2 H_b(\delta) + \log\left( n {\bold{d}_T(n)}\right)}
\end{align}
where $n^2 \alpha_k = k$. Inequality $a$ is due to Stirling and as shown in \cite[Example 11.1.3]{cover}. Inequality $b$ is due to our assumption that $\delta \leq \frac{1}{2}$ and the fact that the summation has {at most} {less than} $n {\bold{d}_T(n)}$ values.
Taking the log on both sides we have
\begin{align} \label{eq:lem:strongTypicalityCardinalityFinal}
- \gamma n \log n \leq \log \left| \rvec{T}_{\delta}( \m{F}_T, {\m{W}} ) \right| - H( \m{F}_T ) \leq n^2 H_b(\delta) + \log n + \log {\bold{d}_T(n)}.
\end{align}
Dividing the above by $n^2$ we obtain the desired result.
\end{IEEEproof}


One last basic claim regarding the $\delta$ edge type classes follows the claim in \cite[Lemma 2.14]{csiszar_korner}. The claim asserts that no ``large probability set'' can be substantially smaller than $\rvec{T}_{\delta}( \m{F}_T, {\m{W}} )$ for $T = \left( \rvecr{r}, \rvecr{c}, \m{W} \right)$.

\begin{lem} \label{lem:lowerBoundedCardinalityOfAnySetOfHighProbability}
{For any arbitrary edge-type $T(\vec{r},\vec{c},\m{W})$ over $[n]$, with density $\bold{d}_T(n)$, if there exists an $\eta \in (0,1)$ complying with the following condition
\begin{align} \label{lem:assumptionOnNwr2eta}
4 n e^{-2 \frac{{\bold{d}_T(n)}^2}{n} \delta^2} \leq \frac{\eta}{2}
\end{align}}
for some small $\delta \in \left[0, \frac{1}{2} \right)$, then if $A$ is some set of graphs over $[n]$ such that its probability according to some $\m{F} \in \mathcal{P}(\vec{r}, \vec{c}, \m{W})$ ($\m{F} {\sim \{ p_{ij} \}}$) is:
\begin{align}
{\Pr}_{\m{F}} \left( A \right) \geq \eta
\end{align}
then we have that
\begin{align}
\frac{1}{n^2} \log | A | \geq  \frac{1}{n^2} H( \m{F}_T ) - H_b(\delta) + \frac{1}{n^2}\log\left( \frac{\eta}{2} \right) - {\frac{\gamma}{n} \log( n)} -{\frac{2}{n} \log ( \bold{d}_T(n) + 1)} - \frac{ \log \left( n{\bold{d}_T(n)} \right) }{n^2}
\end{align}
{where $\gamma > 0$ is the universal constant defined in Theorem \ref{thm:BarvinokThm1}.}
\end{lem}
\begin{IEEEproof}
The proof is given in Appendix \ref{appendix:lem:lowerBoundedCardinalityOfAnySetOfHighProbability}.
\end{IEEEproof}

\section{Conditional Edge-Types} \label{sec:ConditionalEdgeTypes}
Following the {Method of Types} \cite{csiszar_korner} we wish to consider the relationship between two directed graphs, $\m{G}$ and $\m{H}$, over $[n]$ nodes.
For the purpose of exploring the relation between the two graphs we are interested in defining a conditional edge type. To make things clearer, note that we can write the graph $\m{H}$ as follows:
\begin{align} \label{eq:HinTermsOfG}
\m{H} = \m{G} \oplus \m{D}_1 \oplus \m{D}_0
\end{align}
where $\m{D}_1$ is a subgraph of $\m{G}$ and $\m{D}_0$ is a subgraph of $\bar{\m{G}}$ - the complement graph of $\m{G}$. In words, the graph $\m{D}_1$ determines the edges removed from $\m{G}$ and $\m{D}_0$ determines the edges added to $\m{G}$ in the construction of $\m{H}$.
Since $\m{D}_0$ and $\m{D}_1$ are complementary with respect to the graph $\m{G}$, we can look directly at
\begin{align}
\m{D} = \m{D}_1 + \m{D}_0
\end{align}
knowing that for a given graph $\m{G}$ the graph $\m{D}$ defines the edges either removed from $\m{G}$ or added to $\m{G}$. As such, instead of \eqref{eq:HinTermsOfG} we can write $\m{H}$ as follows:
\begin{align} \label{eq:HinTermsOfG2}
\m{H} = \m{G} \oplus \m{D}.
\end{align}
{We can refer to the graph $\m{D}$ as the distortion graph. }

{Given the above discussion and \eqref{eq:HinTermsOfG2} we determine the conditional edge-type class using the base-line graph $\m{G}$ and an edge-type $T = (\vec{r}, \vec{c})$, which determines the possible distortion graphs $\m{D}$ in \eqref{eq:HinTermsOfG2}. The definition is as follows:
\begin{dfn} \label{dfn:conditionalEdgeType}
Let $\m{H}$ and $\m{G}$ be two directed graphs over $[n]$. The conditional edge-type class of $\m{H}$ given $\m{G}$ contains all graphs $\hat{\m{H}}$ such that $\hat{\m{H}} \oplus \m{G}$ belong to the edge type $\rvec{T}( \m{H} \oplus \m{G})$. The conditional edge-type class is denoted
\begin{align}
\rvec{T}( \m{H} | \m{G}).
\end{align}
Alternatively, given an edge-type $T = (\vec{r}, \vec{c})$ and a graph $\m{G}$ over $[n]$, the conditional edge-type class of edge-type $T = (\vec{r}, \vec{c})$ given $\m{G}$ contains all graphs $\hat{\m{H}}$ such that $\hat{\m{H}} \oplus \m{G}$ belong to the edge type $T$. In this case the conditional edge-type class will be denoted as
\begin{align}
\rvec{T}( \vec{r}, \vec{c} | \m{G}).
\end{align}
\end{dfn}}

\begin{rem} \label{rem:restrictedConditionalEdgeType}
When $\m{G}$ is restricted by graph $\m{W}$ and we consider the case in which all graphs $\m{H}$ are restricted by graph $\m{W}$ the extension is direct. The \emph{conditional edge-type {class}} is denoted by either {$\rvec{T}(\m{H} | \m{G}, \m{W})$} or {$\rvec{T}( \vec{r}, \vec{c} | \m{G}, \m{W})$}. Note that if $\m{G}$ is restricted by $\m{W}$ but we consider all possible graphs $\m{H}$ without restriction, we can return to the non-restricted definition of the \emph{conditional edge-type class}. If, however $\m{G}$ is not restricted by graph $\m{W}$ but we do want to consider {a restricted conditional edge-type class} we can take a two step solution:
\begin{align}
\hat{\m{G}} = \m{G} \wedge \m{W}
\end{align}
and consider $\rvec{T}( \vec{r}, \vec{c} | \hat{\m{G}}, \m{W})$.
\end{rem}

{From this point forward we will assume that both the base-line graph $\m{G}$ and the conditional edge-type class are restricted by the same graph $\m{W}$.}

\begin{rem} \label{rem:conditionalEdgeTypes}
Note that unlike the standard {method of types}, the conditional edge-type class is not a subset of an edge-type class. In other words, the set $\rvec{T}(\m{H} | \m{G})$ may contain graphs $\m{H}_1$ and $\m{H}_2$ that do not have the same edge-type. What the two graphs have in common is that they can both be written as follows:
\begin{align}
\m{H}_1 & = \m{G} \oplus \m{D}_{\m{H}_1} \nonumber \\
\m{H}_2 & = \m{G} \oplus \m{D}_{\m{H}_2}
\end{align}
where $\m{D}_{\m{H}_1}, \m{D}_{\m{H}_2} \in \rvec{T}(\m{D})$, where $\m{D}$ is such that $\m{H} = \m{G} \oplus \m{D}$.
{
As an example of the above consider graph $\m{G}$ with two vertices ($n=2$) with the following adjacency matrix:
\begin{align}
\mat{A}_{\m{G}} = \left[ \begin{array}{cc} 1 & 1 \\ 1 & 0\end{array} \right]
\end{align}
and consider the edge-type with both $\vec{r}$ and $\vec{c}$ being $(1,1)$ meaning an edge-type class containing two possible graphs $\m{D}_1$ and $\m{D}_2$ with the following adjacency matrices:
\begin{align}
\begin{array}{cc}
\mat{A}_{\m{D}_1} = \left[
\begin{array}{cc} 1 & 0 \\ 0 & 1 \end{array} \right]  & \textrm{ and } \mat{A}_{\m{D}_2} = \left[
\begin{array}{cc} 0 & 1 \\ 1 & 0 \end{array} \right].
\end{array}
\end{align}
The conditional edge-type class $\rvec{T}(\vec{r}, \vec{c} | \m{G})$ contains also two graphs $\m{H}_1$ and $\m{H}_2$ with the following adjacency matrices:
\begin{align}
\begin{array}{cc}
\mat{A}_{\m{H}_1} = \left[
\begin{array}{cc} 0 & 1 \\ 1 & 1 \end{array} \right]  & \textrm{ and } \mat{A}_{\m{H}_2} = \left[
\begin{array}{cc} 1 & 0 \\ 0 & 0 \end{array} \right].
\end{array}
\end{align}
which clearly do not belong to the same edge-type class. }
\end{rem}

Similar questions like the ones we asked regarding edge-type {classes} are also relevant when we consider conditional edge-type {classes}. Given our definition above (Definition \ref{dfn:conditionalEdgeType}) the extension here is trivial. We give the results as follows:
\begin{cor} \label{cor:probabilityOfrandomGraphInConditionalEdgeType}
Consider the directed graph $\m{G}$ and let it be distorted by an independent {random graph} $\m{F} \sim {\{p_{ij}\}} \in \mathbb{D}(n, \m{W})$. Then for any conditional edge-type class $\rvec{T}( {\vec{r}, \vec{c}} | \m{G}, \m{W})$ we have that {for all $\m{H} \in \rvec{T}( \vec{r}, \vec{c} | \m{G}, \m{W})$}
\begin{align}
{\Pr} \left( \m{H} \Big| \m{G} \right) = e^{-H (\m{F}_T) - \sum_{i,j: \left\{\mat{A}_{\m{W}} \right\}_{i,j} = 1} D( {(p_T)_{i,j}} || {p_{ij}} ) }
\end{align}
where $\m{F}_T \sim ( {(p_T)_{i,j}} )$ is the maximum entropy random graph over the polytop $\mathcal{P}(\vec{r}, \vec{c}, \m{W})$ and assuming $\m{F}_T \ll \m{F}$.
\end{cor}
\begin{IEEEproof}
\begin{align}
{\Pr}_{\m{F}} \left( \m{H} \Big| \m{G} \right) & = {\Pr}_{\m{F}} \left( \m{H} \oplus \m{G} \Big| \m{G} \right) \nonumber \\
& = {\Pr}_{\m{F}} \left( \m{F} = \m{H} \oplus \m{G} \Big| \m{G} \right) \nonumber \\
& = {\Pr}_{\m{F}} \left( \m{F} = \m{H} \oplus \m{G} \right) \nonumber \\
& = e^{-H (\m{F}_T) - \sum_{i,j: \left\{\mat{A}_{\m{W}} \right\}_{i,j} = 1} D( {(p_T)_{i,j}} || {p_{ij}} ) }, \quad \forall \m{H}  \in \rvec{T}({\vec{r}, \vec{c}} | \m{G}, \m{W})
\end{align}
where the third equality is due to the independence of the distortion on the choice of source graph $\m{G}$.
\end{IEEEproof}

The above result is a direct extension of Theorem \ref{thm:probabilityOfrandomGraphInEdgeType}. Note that the probability depends only on $\m{F}_T$ which is the maximum entropy random graph over a polytop {defined by the distortion defining the conditional edge-type class considered}.

Similarly, we also have the following corollary of Theorem \ref{thm:probabilityOfAnEdgeType}
\begin{cor} \label{cor:probabilityOfAconditionalEdgeType}
Consider the directed graph $\m{G}$ and let it be distorted by an independent distortion $\m{F} \sim {\{p_{ij}\}} \in \mathbb{D}(n, \m{W})$. Then for {any} conditional edge-type $\rvec{T}({\vec{r}, \vec{c}} | \m{G}, \m{W})$ and maximum entropy random graph $\m{F}_T$ over the polytop $\mathcal{P}(\vec{r}, \vec{c}, \m{W})$, if $\m{F}_T \ll \m{F}$
\begin{align}
(n)^{-2\gamma(2n)} e^{- \sum_{i,j:\left\{\mat{A}_{\m{W}} \right\}_{i,j}=1} D( {(p_T)_{i,j}} || {p_{ij}} ) } \leq \Pr \left( \m{H} \in \rvec{T}({\vec{r}, \vec{c}} | \m{G}, \m{W}) \Big| \m{G} \right) \leq e^{- \sum_{i,j:\left\{\mat{A}_{\m{W}} \right\}_{i,j}=1} D( {(p_T)_{i,j}} || {p_{ij}} ) }
\end{align}
for some absolute constant $\gamma > 0$.
\end{cor}
As in Corollary \ref{cor:probabilityOfrandomGraphInConditionalEdgeType} the probability of a conditional edge-type class depends only on the maximum entropy random graph over the polytop {defined by the distortion of the conditional edge-type class considered}.

Moving to the cardinality of \emph{conditional edge-types} we have the following corollary of Theorem \ref{thm:BarvinokThm1} and Lemma \ref{lem:BarvinokDual}
\begin{cor} \label{cor:cardinalityOfconditionalEdgeTypes}
Given the directed graphs $\m{G}$ and $\m{H}$, the cardinality of the corresponding \emph{conditional edge-type class}, $\rvec{T}(\m{H} | \m{G}, \m{W})$ is bounded as follows:
\begin{align}
2^{-\gamma n\log{n}}e^{H(\m{F}_T)} \leq |\rvec{T}(\m{H} | \m{G}, \m{W})| \leq e^{H(\m{F}_T)}
\end{align}
where $\m{F}_T$ is the maximum entropy random graph over the polytop {$\mathcal{P}(\vec{r}_{\m{H} \oplus \m{G}}, \vec{c}_{\m{H} \oplus \m{G}}, \m{W})$}
and $\gamma > 0$ is a universal constant.
\end{cor}

\subsection{Delta Conditional Edge-Types}
As in Section \ref{sec:MethodOfTypesExtended} we also require the extension of \emph{conditional edge-type classes} to \emph{strong typicality}, meaning to larger classes that allow small variations in the incoming and outgoing degrees. We basically repeat Definition \ref{dfn:strongTypicalityEdgeType} and the following lemmas within the setting of \emph{conditional edge-type classes}.

\begin{dfn} \label{dfn:strongTypicalityConditional}
We say that a graph $\m{H}$ over $[n]$ has $\delta$ edge-type $T = (\rvecr{r}, \rvecr{c}, \m{W})$ with respect to graph $\m{G}$ if both are restricted by graph $\m{W}$ and 
\begin{align}
\frac{1}{{\bold{d}_T(n)}} \left| \rvecr{r}_{\m{H} \oplus \m{G}}(i) - \rvecr{r}(i) \right| & < \delta \quad \forall i \in [n] \nonumber \\
\frac{1}{{\bold{d}_T(n)}} \left| \rvecr{c}_{\m{H} \oplus \m{G}}(i) - \rvecr{c}(i) \right| & < \delta \quad \forall i \in [n].
\end{align}
The set of such graphs is denoted as {either} $\rvec{T}_{\delta}(\m{H} | \m{G}, \m{W})$ {when a member $\m{H}$ is provided or $\rvec{T}_{\delta}(\vec{r}, \vec{c} | \m{G}, \m{W})$ when the pair $(\vec{r}, \vec{c})$ of the distortion is provided}. {Another notation we will be using is $\rvec{T}_{\delta}( \m{F}_T | \m{G}, \m{W} )$ where $\m{F}_T$ is the maximum entropy random graph over $\mathcal{P}(\vec{r}, \vec{c}, \m{W})$.}
\end{dfn}

Extending Lemma \ref{lem:extensionLemma2.12_CK}:
\begin{cor} \label{cor:extensionLemma2.12_CK_conditional}
For any $n$, every $\rvec{T}_{\delta}({\vec{r}, \vec{c}} | \m{G}, \m{W})$ for some small $\delta > 0$ we have that
\begin{align}
\Pr \left( \rvec{T}_{\delta}( {\vec{r}, \vec{c}} | \m{G}, \m{W}) \Big| \m{G} \right) \geq 1 - 4 n e^{-2 {\frac{\bold{d}_T(n)^2}{n} }\delta^2} \nonumber
\end{align}
{where the probability can be according to any random graph from the polytop $\mathcal{P}(\vec{r}, \vec{c}, \m{W})$ independent of $\m{G}$, and $T = (\vec{r}, \vec{c}, \m{W})$.}
\end{cor}

Extending Lemma \ref{lem:strongTypicalityCardinality}
\begin{cor} \label{cor:strongTypicalityCardinalityConditional}
For a given $T = \left( \rvecr{r}, \rvecr{c}, \m{W} \right)$, the cardinality of $\rvec{T}_{\delta}( \rvecr{r}, \rvecr{c} | \m{G}, \m{W} )$ with constant $\delta  \in \left[0, \frac{1}{2} \right)$ {and the universal constant $\gamma > 0$ defined in Theorem \ref{thm:BarvinokThm1},} has the following property:
\begin{align}
- \frac{\gamma \log n}{n} \leq \frac{1}{n^2}\log \left| \rvec{T}_{\delta}( \rvecr{r}, \rvecr{c} | \m{G}, \m{W} ) \right| - \frac{1}{n^2} H( \m{F}_T ) \leq H_b(\delta) + \frac{\log n {\bold{d}_T(n)}}{n^2}.
\end{align}
\end{cor}

Extending Lemma \ref{lem:lowerBoundedCardinalityOfAnySetOfHighProbability}
\begin{cor} \label{lem:lowerBoundedCardinalityOfAnySetOfHighProbabilityConditional}
{For any arbitrary edge-type $T(\vec{r},\vec{c},\m{W})$ over $[n]$, with density $\bold{d}_T(n)$, if there exists an $\eta \in (0,1)$ complying with the following condition
\begin{align} \label{lem:assumptionOnNwr2eta_conditional}
4 n e^{-2 \frac{{\bold{d}_T(n)}^2}{n} \delta^2} \leq \frac{\eta}{2}
\end{align}}
{for some small $\delta \in \left[0, \frac{1}{2} \right)$ then, if $A$ is some set of graphs over $[n]$ such that the probability of the set $\{ \m{H} \oplus \m{G}: \m{H} \in A \}$ according to some $\m{F} \in \mathcal{P}(\vec{r},\vec{c}, \m{W})$ ($\m{F} \sim \{ p_{ij} \}$) is:}
\begin{align}
{\Pr} \left( \{ \m{H} \oplus \m{G}: \m{H} \in A \} \right) \geq \eta
\end{align}
then we have that
\begin{align}
\frac{1}{n^2} \log | A | \geq  \frac{1}{n^2} H( \m{F}_T ) - H_b(\delta) + \frac{1}{n^2}\log\left( \frac{\eta}{2} \right) - {\frac{\gamma}{n} \log( n)} -{\frac{2}{n} \log(\bold{d}_T(n)+1)}  - \frac{ \log \left( n{\bold{d}_T(n)} \right) }{n^2}.
\end{align}
\end{cor}

\section{Rate-Distortion} \label{sec:RateDistortionGraphs}

In this section we first prove an equivalent result to the \emph{Covering Lemma} which sharpens the achievability part of the rate-distortion theorem {(Theorem \ref{thm:0fidelity_rateDistortion} given next)}. {In a nutshell, the Covering Lemma considers an edge-type class of graphs and a distortion constraint (assuming the distortion measure defined in \eqref{eq:distortion_edge_type}). For this edge-type class the Covering Lemma claims the existence of a set of graphs (denoted as $B$) that can act as \emph{representatives}, meaning that the distortion constraint of any graph in the edge-type class is held with respect to at least one of the \emph{representatives} in $B$. Surely such a set $B$ exists, as we can always take the entire edge-type class as a set of \emph{representatives}, so the added value of the claim is that it bounds the cardinality of this set and shows that a smaller cardinality set exists that provides this property.}

{The proof follows the ideas (and also partially the notation) presented in the proof of the covering lemma provided in} \cite[Lemma 2.4.1]{csiszar_korner}.

\begin{lem} \label{lem:coveringLemma}
For distortion measure $d(\cdot, \cdot)$ as defined in \eqref{eq:distortion_edge_type}, an edge-type class $\rvec{T}( \rvecr{r}, \rvecr{c}, \m{W})$ {over $[n]$ and density $\bold{d}(n)$,} and number $\Xi \geq 0$ there exists a set $B$ of directed graphs over $[n]$ such that
\begin{align}
d( \m{G}, B ) \equiv \min_{\m{H} \in B} d( \m{G}, \m{H} ) \leq \Xi {+ \frac{\delta}{n}}, \quad \textrm{ for every } \m{G} \in \rvec{T}(\rvecr{r}, \rvecr{c}, \m{W})
\end{align}
{for some small $\delta >0$,}
and
\begin{align}
\frac{1}{n^2} \log \big| B \big| & \leq \max_{(\rvecr{d}_r, \rvecr{d}_c) \in \Omega(\Xi)}  \frac{1}{n^2} \left\{ H( \m{F}_{\rvecr{r} \pm \rvecr{d}_r, \rvecr{c} \pm \rvecr{d}_c , \m{W}}  ) -H( \m{F}_{\rvecr{d}_r, \rvecr{d}_c,\m{W}} ) \right\} \nonumber \\
& + \frac{ (2 \Xi {n} + 2)}{n^2} \log n + H_b( \delta) + {\frac{1}{n^2} \log\left( n \bold{d}(n) \right)} + \frac{\gamma}{{n}} \log n + \frac{1}{n}
\end{align}
for large enough\footnote{$n$ has to be large enough such that $e^{n^2 - e^{n}} < 1$.} $n$, where
\begin{align} \label{eq:Sigma}
\Omega( \Xi ) = \left\{\left( \rvecr{d}_r, \rvecr{d}_c \right):  {\frac{1}{n}} \max\left\{ {\|\rvecr{d}_r\|_\infty \,,\, \|\rvecr{d}_c\|_\infty}\right\} \leq \Xi, \rvecr{d}_r,\rvecr{d}_c \geq 0 \right\}
\end{align}
{where by $\pm$ we mean that each element in the vector $\rvecr{r}$ ($\rvecr{c}$) can either increase by the corresponding element in $\rvecr{d}_r$ ($\rvecr{d}_c$) or decrease by the corresponding element in $\rvecr{d}_r$ ($\rvecr{d}_c$). Moreover, we assume that the distortion does not change the density property, meaning:
\begin{align} \label{eq:assumptionFixedDensity_CoveringLemma}
\bold{d}_{\rvec{T}( \vec{r} \pm \rvecr{d}_{\vec{r}}, \vec{c} \pm \rvecr{d}_{\vec{c}}, \m{W})}(n) = \bold{d}(n).
\end{align}
}
\end{lem}


In the proof we follow the proof of \cite{csiszar_korner} but with a slight difference. The set from which we construct the covering set is considerably larger, and contains several types. However, we can bound the number of these types and in exponential terms there are $e^{2 n \log n}$ such types. This allows us to continue following the proof by simply considering the cardinality of the largest among these types.

\begin{IEEEproof}
The basic idea in the proof of the covering lemma as appears in \cite[Lemma 2.4.1]{csiszar_korner} is \emph{random selection}. We want to show the \emph{existence} of a covering set without specifying it precisely. This is done by examining the set of graphs (from the given edge-type class, $\rvec{T}( \rvecr{r}, \rvecr{c}, \m{W})$) not covered by a given covering set. Now, showing that the cardinality of such sets, of graphs not covered, is smaller than one in expectation (over all possible covering sets) guarantees that there exists at least one such covering set that covers all graphs, and for that covering set the set of graphs not covered is an empty set.

The main difference form the covering lemma in \cite[Lemma 2.4.1]{csiszar_korner} is that the covering sets are taken over a more complex set. As mentioned in Remark \ref{rem:conditionalEdgeTypes}, two graphs $\m{H}_1$ and $\m{H}_2$ of a given conditional edge-type class are not necessarily of the same edge-type. Therefore, we consider a union of {conditional $\delta$} edge-type classes as the source of our covering sets. {Specifically,}
\begin{align} \label{eq:unionForSetB_first}
B \subset \cup_{ \m{G} \in \rvec{T}( \rvecr{r}, \rvecr{c}, \m{W})} \cup_{ (\rvecr{d}_r, \rvecr{d}_c) \in \Omega( \Xi)} {\rvec{T}_{\delta}( \rvecr{d}_r, \rvecr{d}_c | \m{G}, \m{W})}.
\end{align}
The above set is a union of {conditional $\delta$} edge-type classes which {considerably overlap one another.} {Consider, for simplicity, the above as a union of conditional edge-types (as apposed to conditional $\delta$ edge-types),} two {matrices} {graphs} $\m{G}_1$ and $\m{G}_2$ from $\rvec{T}( \rvecr{r}, \rvecr{c}, \m{W})$ with different distortions can result {in} the same {graph}.
{As such,} we define a different set, {this time as a union of $\delta$ edge-types and not conditional $\delta$ edge-types}, {in} which {the overlapping is due only to the $\delta$ perturbations.} As will be evident, every subset $B$ in \eqref{eq:unionForSetB_first} is also a subset of this new union, as this new union contains the union in \eqref{eq:unionForSetB_first}:
\begin{align} \label{eq:unionForSetB_second}
B \subset \cup_{(\rvecr{d}_r, \rvecr{d}_c) \in \Omega(\Xi)} \rvec{T}_{\delta}( \rvecr{r} \pm \rvecr{d}_r, \rvecr{c} \pm \rvecr{d}_c, \m{W})
\end{align}
where by $\pm$ we mean that each element in the vector $\rvecr{r}$ ($\rvecr{c}$) can either be increase by the corresponding element in $\rvecr{d}_r$ ($\rvecr{d}_c$) or decrease by the corresponding element in $\rvecr{d}_r$ ($\rvecr{d}_c$). In other words, we have that a given pair, $\left( \rvecr{d}_r, \rvecr{d}_c \right)$, produces at most $2^{2 n}$ possible pairs, determining the edge-type. Surely not all options are valid ones and not all produce different vectors (\eg, if an element is zero there is no distinction between increasing or reducing by its value), thus {$2^{2 n}$} provides only a {crude} upper bound.

{By definition of conditional $\delta$ edge-types}, every graph $\m{H}$ in the union presented in \eqref{eq:unionForSetB_first} {can be written as $\m{G} \oplus \m{D}$, for some $\m{G} \in \rvec{T}(\vec{r}, \vec{c}, \m{W})$ and some $\m{D} \in \rvec{T}_{\delta}( \rvecr{d}_r, \rvecr{d}_c, \m{W})$ for some pair $(\rvecr{d}_r, \rvecr{d}_c) \in \Omega(\Xi)$. 
}
This means that the graph $\m{H}$ is limited to one of {the $\delta$ edge-types of the form} $\rvec{T}_{\delta}( \rvecr{r} \pm \rvecr{d}_r, \rvecr{c} \pm \rvecr{d}_c, \m{W})$ where $(\rvecr{d}_r, \rvecr{d}_c)$ are limited according to $\Omega(\Xi)$ \eqref{eq:Sigma}. Note that the other direction does not necessarily hold.

Following \eqref{eq:unionForSetB_second} we have a union of {conditional $\delta$} edge-type classes. Thus, the cardinality {can be upper bounded by} the sum {of cardinalities}. We first bound the cardinality of $\Omega(\Xi)$ to obtain an upper bound on the number of pairs $( \rvecr{d}_r, \rvecr{d}_c)$. We have that
\begin{align} \label{eq:boundOnOmega}
\left| \Omega(\Xi) \right| \leq n^{\Xi {n} + 1} n^{\Xi {n} + 1} {= e^{2( \Xi n + 1) \log n}}
\end{align}
since either $\rvecr{d}_r$ or $\rvecr{d}_c$ can each receive any value in the range $[0, \Xi {n}]$ ($\Xi {n} + 1$ values) for each of their $n$ components.
Using this bound and the fact that {the overlap between the conditional $\delta$ edge-types is only due to the $\delta$ perturbation} we will be able to provide {a} tight bound on the entire union.

We denote the following set
\begin{align}
U( B ) = \{ \m{G} | \m{G} \in \rvec{T}( \rvecr{r}, \rvecr{c}, \m{W}), d( \m{G}, B) > \Xi + \frac{\delta}{n}\}
\end{align}
{which is the set of all graphs $\m{G}$ that are not properly represented, with respect to the distortion measure and its constraint, by a given subset $B$.}

Fix $M > 0$ and {let} $\mathcal{B}^M$ be the family of all collections of {at most} $M$ elements in $\cup_{(\rvecr{d}_r, \rvecr{d}_c) \in \Omega(\Xi)} \rvec{T}_{\delta}( \rvecr{r} \pm \rvecr{d}_r, \rvecr{c} \pm \rvecr{d}_c, \m{W})$, in other words, all possible {sub}sets of cardinality {bounded by} $M$. We also define $\rvec{Z}^M$ to be a random variable ranging over $\mathcal{B}^M$ with uniform distribution. Thus,
\begin{align}
\rvec{Z}^M = \left( \mat{Z}_1, \mat{Z}_2, \ldots, \mat{Z}_M \right)
\end{align}
where $\mat{Z}_i$ are independent and uniformly distributed over $\cup_{(\rvecr{d}_r, \rvecr{d}_c) \in \Omega(\Xi)} \rvec{T}_{\delta}( \rvecr{r} \pm \rvecr{d}_r, \rvecr{c} \pm \rvecr{d}_c, \m{W})$. {Note that since $\mat{Z}_i$ are independent a realization of $\rvec{Z}^M$ may be a subset containing identical elements from $\cup_{(\rvecr{d}_r, \rvecr{d}_c) \in \Omega(\Xi)} \rvec{T}_{\delta}( \rvecr{r} \pm \rvecr{d}_r, \rvecr{c} \pm \rvecr{d}_c, \m{W})$ in which case the cardinality of the specific realization is less than $M$.}

Now, we consider $\Esp{ \left| U ( \rvec{Z}^M ) \right| }$ and our goal is to show that it is strictly bounded from above by one, meaning that in average over all possible covering sets the cardinality of \emph{un-covered} inputs is smaller than one. From this we can conclude the \emph{existence} of at least one \emph{good} covering set, for which the set of \emph{un-covered} inputs is empty.

We denote by $\chi( \m{G} )$ the characteristic function of the random set $U( \rvec{Z}^M)$
\begin{align}
\chi( \m{G} ) = \left\{ \begin{array}{cc} 1 & \textrm{if } \m{G} \in U( \rvec{Z}^M) \\ 0 & \textrm{if } \m{G} \notin U( \rvec{Z}^M) \end{array} \right.
\end{align}
Then
\begin{align}
\left| U ( \rvec{Z}^M ) \right| & = \sum_{\m{G} \in \rvec{T}(\rvecr{r}, \rvecr{c}, \m{W})} \chi( \m{G} ) \nonumber \\
\Esp{\left| U ( \rvec{Z}^M ) \right| } & = \sum_{\m{G} \in \rvec{T}( \rvecr{r}, \rvecr{c}, \m{W})} \Esp{ \chi( \m{G} ) } = \sum_{\m{G} \in \rvec{T}(\rvecr{r}, \rvecr{c}, \m{W})} \Pr \left( \m{G} \in U ( \rvec{Z}^M ) \right). \label{eq:rateDistortion:expectationOfU}
\end{align}

We examine the probability that a specific graph $\m{G} \in \rvec{T}(\vec{r}, \vec{c}, \m{W})$ does not comply with the distortion constraint. We begin by examining the complementary probability:
\begin{align}
\Pr \left\{ d( \m{G}, \rvec{Z}_i ) > \Xi  + \frac{\delta}{n} \right\} & = 1 - \Pr \left\{ d( \m{G}, \rvec{Z}_i ) \leq \Xi  + \frac{\delta}{n} \right\}.
\end{align}
{We first claim that for any $\m{H} \in \cup_{(\rvecr{d}_r, \rvecr{d}_c) \in \Omega(\Xi)} \rvec{T}_{\delta}(\rvecr{d}_r, \rvecr{d}_c | \m{G}, \m{W})$
\begin{align}
d( \m{G}, \m{H}) \leq \Xi + \frac{\delta}{n}.
\end{align}
This is shown precisely in Appendix \ref{appendix:transitionToDeltaEdgeTypeInCoveringLemma}.}
We now lower bound the complementary probability by noting that it is at least the fraction of graphs matched to $\m{G}$ {in the sense that their distortion with respect to $\m{G}$ complies with the constraint. These are the graphs belonging to $\cup_{(\rvecr{d}_r, \rvecr{d}_c) \in \Omega(\Xi)} \rvec{T}_{\delta}(\rvecr{d}_r, \rvecr{d}_c | \m{G}, \m{W})$, as shown above}:
\begin{align}
\Pr \left\{ d( \m{G}, \rvec{Z}_i ) \leq \Xi  + \frac{\delta}{n} \right\} & \geq \frac{ \left| \cup_{(\rvecr{d}_r, \rvecr{d}_c) \in \Omega(\Xi) } {\rvec{T}_{\delta} ( \rvecr{d}_r, \rvecr{d}_c |\m{G}, \m{W})} \right| }{\left| \cup_{(\rvecr{d}_r, \rvecr{d}_c) \in \Omega(\Xi)}  \rvec{T}_{{\delta}}( \rvecr{r} \pm \rvecr{d}_r, \rvecr{c} \pm \rvecr{d}_c , \m{W}) \right|  } \nonumber \\
& {\geq \frac{ \left| \cup_{(\rvecr{d}_r, \rvecr{d}_c) \in \Omega(\Xi) } {\rvec{T}_{\delta} ( \rvecr{d}_r, \rvecr{d}_c |\m{G}, \m{W})} \right| }{\sum_{(\rvecr{d}_r, \rvecr{d}_c) \in \Omega(\Xi)} \left| \rvec{T}_{{\delta}}( \rvecr{r} \pm \rvecr{d}_r, \rvecr{c} \pm \rvecr{d}_c , \m{W}) \right|  }}.
\end{align}
{The second inequality is due to the fact that the $\delta$ edge types might overlap due to the $\delta$ perturbations. Thus, by taking the summation over their cardinality we are increasing the denominator.}
Putting the two together we have that
\begin{align}
\Pr \left\{ d( \m{G}, \rvec{Z}_i ) > \Xi  + \frac{\delta}{n} \right\} & \leq 1 - \frac{ \left| \cup_{(\rvecr{d}_r, \rvecr{d}_c) \in \Omega(\Xi) } {\rvec{T}_{\delta} ( \rvecr{d}_r, \rvecr{d}_c |\m{G}, \m{W})} \right| }{{\sum_{(\rvecr{d}_r, \rvecr{d}_c) \in \Omega(\Xi)}} \left| \rvec{T}_{{\delta}}( \rvecr{r} \pm \rvecr{d}_r, \rvecr{c} \pm \rvecr{d}_c , \m{W}) \right|  } \nonumber \\
& \stackrel{a}{\leq} 1 - \frac{{ \max_{(\rvecr{d}_r, \rvecr{d}_c) \in \Omega(\Xi)}} \left|  \rvec{T}_{{\delta}} ( { {\rvecr{d}}_r, {{\rvecr{d}}}_c} |\m{G}, \m{W})  \right| }{\sum_{(\rvecr{d}_r, \rvecr{d}_c) \in \Omega(\Xi)}  \left| \rvec{T}_{{\delta}}( \rvecr{r} \pm \rvecr{d}_r, \rvecr{c} \pm \rvecr{d}_c , \m{W}) \right|  } \nonumber \\
& \stackrel{b}{\leq} 1 - \frac{ { \max_{(\rvecr{d}_r, \rvecr{d}_c) \in \Omega(\Xi)}} \left|  \rvec{T}_{{\delta}} ( { {\rvecr{d}}_r, {\rvecr{d}}_c } |\m{G}, \m{W})  \right| }{ e^{(2\Xi {n} + 2) \log n} \max_{(\rvecr{d}_r, \rvecr{d}_c) \in \Omega(\Xi)}  \left| \rvec{T}_{{\delta}}( \rvecr{r} \pm \rvecr{d}_r, \rvecr{c} \pm \rvecr{d}_c , \m{W}) \right|  } \nonumber \\
& \stackrel{c}{\leq} 1 - \frac{ { \max_{(\rvecr{d}_r, \rvecr{d}_c) \in \Omega(\Xi)}} e^{H( \m{F}_{{{\rvecr{d}}_r, {\rvecr{d}}_c},\m{W}} ) - \gamma n \log n}  }{ e^{(2\Xi{n} + 2) \log n} \max_{(\rvecr{d}_r, \rvecr{d}_c) \in \Omega(\Xi)}  e^{ H( \m{F}_{\rvecr{r} \pm \rvecr{d}_r, \rvecr{c} \pm \rvecr{d}_c , \m{W}}  ) + n^2 H_b( \delta) +  {\log\left( n \bold{d}_{\rvecr{r} \pm \rvecr{d}_r, \rvecr{c} \pm \rvecr{d}_c}(n) \right)} } } \nonumber \\
& = 1 - \left( \frac{ e^{(2\Xi{n} + 2) \log n} \max_{(\rvecr{d}_r, \rvecr{d}_c) \in \Omega(\Xi)}  e^{ H( \m{F}_{\rvecr{r} \pm \rvecr{d}_r, \rvecr{c} \pm \rvecr{d}_c , \m{W}}  ) + n^2 H_b( \delta) + {\log\left( n \bold{d}_{\rvecr{r} \pm \rvecr{d}_r, \rvecr{c} \pm \rvecr{d}_c}(n) \right)} } }{ { \max_{(\rvecr{d}_r, \rvecr{d}_c) \in \Omega(\Xi)}} e^{H( \m{F}_{{ {\rvecr{d}}_r, {\rvecr{d}}_c},\m{W}} ) - \gamma n \log n}  }  \right)^{-1} \nonumber \\
& \stackrel{d}{{\leq}} 1 - \left(    e^{ \max_{(\rvecr{d}_r, \rvecr{d}_c) \in \Omega(\Xi)}  \left\{ H( \m{F}_{\rvecr{r} \pm \rvecr{d}_r, \rvecr{c} \pm \rvecr{d}_c , \m{W}}  ) -H( \m{F}_{\rvecr{d}_r, \rvecr{d}_c,\m{W}} ) \right\} +  ( 2 \Xi {n} + 2) \log n + n^2 H_b( \delta) + {\log\left( n \bold{d}(n) \right)} + \gamma n \log n}   \right)^{-1} \nonumber \\
& = 1 -  e^{ - \max_{(\rvecr{d}_r, \rvecr{d}_c) \in \Omega(\Xi)}  \left\{ H( \m{F}_{\rvecr{r} \pm \rvecr{d}_r, \rvecr{c} \pm \rvecr{d}_c , \m{W}}  ) -H( \m{F}_{\rvecr{d}_r, \rvecr{d}_c,\m{W}} ) \right\} -  ( 2 \Xi {n} + 2) \log n - n^2 H_b( \delta) - {\log\left( n \bold{d}(n) \right)} - \gamma n \log n}
\end{align}
where the (in)equalities are due to:
\begin{enumerate}
\item $a$ - choosing a specific pair ${({\rvecr{d}}_r, {\rvecr{d}}_c)}$ from the union. This bound holds for any such pair. {We specifically choose the maximizing pair.}
\item $b$ - increasing the denominator to the maximum size according to the cardinality of $\Omega(\Xi)$ and the maximum cardinality term in the summation.
\item $c$ - taking the corresponding bounds on the cardinality. {The lower bound on the cardinality of conditional} $\delta$ edge-types for the nominator (Corollary \ref{cor:strongTypicalityCardinalityConditional}), and the upper bound of $\delta$ edge types for the denominator (Lemma \ref{lem:strongTypicalityCardinality}).
\item $d$ - {since the bound holds for any arbitrary pair $(\hat{\rvecr{d}}_r, \hat{\rvecr{d}}_c)$ we can place both into the same maximization.} {this transition follows due to:
    \begin{align}
    & \max_{(\rvecr{d}_r, \rvecr{d}_c) \in \Omega(\Xi)} H( \m{F}_{\rvecr{r} \pm \rvecr{d}_r, \rvecr{c} \pm \rvecr{d}_c , \m{W}}  ) - \max_{(\rvecr{d}_r, \rvecr{d}_c) \in \Omega(\Xi)} H( \m{F}_{\rvecr{d}_r, \rvecr{d}_c,\m{W}} ) \nonumber \\
    & = \max_{(\rvecr{d}_r, \rvecr{d}_c) \in \Omega(\Xi)} H( \m{F}_{\rvecr{r} \pm \rvecr{d}_r, \rvecr{c} \pm \rvecr{d}_c , \m{W}}  ) + \min_{(\rvecr{d}_r, \rvecr{d}_c) \in \Omega(\Xi)} - H( \m{F}_{\rvecr{d}_r, \rvecr{d}_c,\m{W}} ) \nonumber \\
    & \leq \max_{(\rvecr{d}_r, \rvecr{d}_c) \in \Omega(\Xi)} \left\{ H( \m{F}_{\rvecr{r} \pm \rvecr{d}_r, \rvecr{c} \pm \rvecr{d}_c , \m{W}}  ) - H( \m{F}_{\rvecr{d}_r, \rvecr{d}_c,\m{W}} ) \right\}
    \end{align}
    where the inequality is due to the fact that if
    \begin{align}
    (\rvecr{d}^\star_r, \rvecr{d}^\star_c) = \textrm{arg} \max_{(\rvecr{d}_r, \rvecr{d}_c) \in \Omega(\Xi)} H( \m{F}_{\rvecr{r} \pm \rvecr{d}_r, \rvecr{c} \pm \rvecr{d}_c , \m{W}}  )
    \end{align}
    then
    \begin{align}
    & \max_{(\rvecr{d}_r, \rvecr{d}_c) \in \Omega(\Xi)} H( \m{F}_{\rvecr{r} \pm \rvecr{d}_r, \rvecr{c} \pm \rvecr{d}_c , \m{W}}  ) + \min_{(\rvecr{d}_r, \rvecr{d}_c) \in \Omega(\Xi)} - H( \m{F}_{\rvecr{d}_r, \rvecr{d}_c,\m{W}} ) \nonumber \\
    & \leq
    H( \m{F}_{\rvecr{r} \pm \rvecr{d}^\star_r, \rvecr{c} \pm \rvecr{d}^\star_c , \m{W}}  ) - H( \m{F}_{\rvecr{d}^\star_r, \rvecr{d}^\star_c,\m{W}} ) \nonumber \\
    & \leq \max_{(\rvecr{d}_r, \rvecr{d}_c) \in \Omega(\Xi)} \left\{H( \m{F}_{\rvecr{r} \pm \rvecr{d}_r, \rvecr{c} \pm \rvecr{d}_c , \m{W}}  ) - H( \m{F}_{\rvecr{d}_r, \rvecr{d}_c,\m{W}} ) \right\}.
    \end{align}
    } In this transition we also use the assumption on the effect of the distortion on the density as given in \eqref{eq:assumptionFixedDensity_CoveringLemma}.
\end{enumerate}
Note that as compared to the standard \emph{covering lemma} (see \cite[Lemma 2.4.1]{csiszar_korner}) the entire set from which we construct $B$ is considerably larger and contains several different edge types. However, the number of edge types composing this larger set is bounded and is at most $e^{(2 \Xi {n} + 2) \log n}$.

Using the above we have that
\begin{align}
\Pr \left\{ \m{G} \in U( \rvec{Z}^M ) \right\} & = \Pr \left\{ d( \m{G}, \rvec{Z}_i) > \Xi + \frac{\delta}{n}, \forall i\in [1,M] \right\} \nonumber \\
& \stackrel{a}{=} \prod_{i=1}^M \Pr \left\{ d( \m{G}, \rvec{Z}_i) > \Xi + \frac{\delta}{n} \right\} \nonumber \\
& \leq \left[ 1 - e^{ - \max_{(\rvecr{d}_r, \rvecr{d}_c) \in \Omega(\Xi)}  \left\{ H( \m{F}_{\rvecr{r} \pm \rvecr{d}_r, \rvecr{c} \pm \rvecr{d}_c , \m{W}}  ) -H( \m{F}_{\rvecr{d}_r, \rvecr{d}_c,\m{W}} ) \right\} - ( 2 \Xi {n} + 2) \log n - n^2 H_b( \delta) - {\log\left( n \bold{d}(n) \right)} - \gamma n \log n} \right]^M \nonumber \\
\end{align}
where transition $a$ is due to the independence of the $\rvec{Z}_i$s $i \in [1,M]$.

Following the proof in \cite[Lemma 1.4.1]{csiszar_korner} we use the inequality $(1-t)^M \leq e^{-tM}$ and obtain
\begin{align}
\Pr \left\{ \m{G} \in U( \rvec{Z}^M ) \right\} & \leq \left[ 1 - e^{ - \max_{(\rvecr{d}_r, \rvecr{d}_c) \in \Omega(\Xi)}  \left\{ H( \m{F}_{\rvecr{r} \pm \rvecr{d}_r, \rvecr{c} \pm \rvecr{d}_c , \m{W}}  ) -H( \m{F}_{\rvecr{d}_r, \rvecr{d}_c,\m{W}} ) \right\} - ( 2 \Xi {n} + 2) \log n - n^2 H_b( \delta) - {\log\left( n \bold{d}(n) \right)} - \gamma n log n} \right]^M \nonumber \\
& \leq \textrm{exp}\left\{-M e^{ - \max_{(\rvecr{d}_r, \rvecr{d}_c) \in \Omega(\Xi)}  \left\{ H( \m{F}_{\rvecr{r} \pm \rvecr{d}_r, \rvecr{c} \pm \rvecr{d}_c , \m{W}}  ) -H( \m{F}_{\rvecr{d}_r, \rvecr{d}_c,\m{W}} ) \right\} - ( 2 \Xi {n} + 2) \log n - n^2 H_b( \delta) - {\log\left( n \bold{d}(n) \right)} - \gamma n \log n } \right\}
\end{align}
Setting
\begin{align}
M = e^{ \max_{(\rvecr{d}_r, \rvecr{d}_c) \in \Omega(\Xi)}  \left\{ H( \m{F}_{\rvecr{r} \pm \rvecr{d}_r, \rvecr{c} \pm \rvecr{d}_c , \m{W}}  ) -H( \m{F}_{\rvecr{d}_r, \rvecr{d}_c,\m{W}} ) \right\} + ( 2 \Xi {n} + 2) \log n + n^2 H_b( \delta) + {\log\left( n \bold{d}(n) \right)} + \gamma n \log n + n }
\end{align}
{with the added $+n$ in the exponent} we have that
\begin{align}
\Pr \left\{ \m{G} \in U( \rvec{Z}^M ) \right\} & \leq \textrm{exp}\left\{-e^{n}\right\}.
\end{align}
Returning to the target of the calculation:
\begin{align}
\Esp{ \left| U( \rvec{Z}^M ) \right| } & \leq \left| \rvec{T}( \rvecr{r}, \rvecr{c}, \m{W} ) \right| \Pr \left\{ \m{G} \in U( \rvec{Z}^M ) \right\} \nonumber \\
& \leq \textrm{exp}\left\{H\left( \m{F}_{\rvecr{r}, \rvecr{c}, \m{W}} \right) - e^{n}\right\} \nonumber \\
& \leq \textrm{exp}\left\{n^2 - e^{n}\right\} \nonumber \\
& < 1
\end{align}
for large enough values of $n$. This proves the existence of a covering set of size $M$. Thus,
\begin{multline}
\frac{1}{n^2} \log M  = \\  \max_{(\rvecr{d}_r, \rvecr{d}_c) \in \Omega(\Xi)}  \frac{1}{n^2} \left\{ H( \m{F}_{\rvecr{r} \pm \rvecr{d}_r, \rvecr{c} \pm \rvecr{d}_c , \m{W}}  ) -H( \m{F}_{\rvecr{d}_r, \rvecr{d}_c,\m{W}} ) \right\} + \frac{ 2 \Xi {n} + 2}{n^2} \log n + H_b( \delta) + {\frac{1}{n^2} \log\left( n \bold{d}(n) \right)} + \frac{\gamma}{{n}} \log n + \frac{1}{n}
\end{multline}
This concludes the proof.
\end{IEEEproof}

Our main result regards the compression of directed graphs given the distortion measure \eqref{eq:distortion_edge_type} which preserves the local structure of the graph. The result provides a lower and upper bound on the combinatorial rate-distortion function, under some assumptions. The \emph{Covering Lemma} established above is used to obtain the achievability result, meaning the upper bound in the next result.

\begin{thm} \label{thm:0fidelity_rateDistortion}
{
Consider an arbitrary edge-type class over $[n]$, $\rvec{T}(\vec{r}, \vec{c}, \m{W})$, with density $\bold{d}(n)$, such that
\begin{align} \label{eq:assumptionNandD4usingCardinalityOfHighProbabilitySet_rateDistrotion}
4 n e^{-2 \frac{{\bold{d}(n)}^2}{n} \hat{\delta}^2} < \frac{1}{2}
\end{align}
for some small $\hat{\delta} \in \left[0, \frac{1}{2}\right)$.
For every $\Xi > 0$, and some small $\delta > 0$, if for all $(\rvecr{d}_{\vec{r}}, \rvecr{d}_{\vec{c}}) \in \Omega(\Xi)$ we have that
\begin{align} \label{eq:assumptionFixedDensity_rateDistortion}
\bold{d}_{\rvec{T}( \vec{r} \pm \rvecr{d}_{\vec{r}}, \vec{c} \pm \rvecr{d}_{\vec{c}}, \m{W})}(n) = \bold{d}(n),
\end{align}
meaning the distortion does not change the density property, we have the following lower and upper bounds on the combinatorial rate distortion function (see Definition \ref{dfn:CombinatorialRateDistortion}), assuming the distortion measure is the one given in \eqref{eq:distortion_edge_type}, and all graphs are taken from the set $\mathcal{S} = \rvec{T}(\vec{r}, \vec{c}, \m{W})$. The upper bound:
\begin{align}
R_n\left(\Xi + \frac{\delta}{n}\right) \leq & \max_{(\rvecr{d}_r, \rvecr{d}_c) \in \Omega(\Xi)}  \frac{1}{n^2} \left\{ H( \m{F}_{\rvecr{r} \pm \rvecr{d}_r, \rvecr{c} \pm \rvecr{d}_c , \m{W}}  ) -H( \m{F}_{\rvecr{d}_r, \rvecr{d}_c,\m{W}} ) \right\} \nonumber \\
& + \frac{ (2 \Xi {n} + 2)}{n^2} \log n + H_b( {\delta}) + {\frac{1}{n^2} \log\left( n \bold{d}(n) \right)} + \frac{\gamma}{{n}} \log n + \frac{1}{n}.
\end{align}
And lower bound:
\begin{align}
R_n\left( \Xi + \frac{\delta}{n} \right) \geq &  \min_{(\rvecr{d}_r, \rvecr{d}_c) \in \Omega( \Xi)} \frac{1}{n^2} \left\{ {H( \m{F}_{\rvec{T}(\vec{r}, \vec{c}, \m{W})} )} - H\left( \m{F}_{ \rvec{T}( \rvecr{d}_r, \rvecr{d}_c, \m{W})} \right) \right\} {- \left( H_b(\hat{\delta}) + H_b({\delta}) \right) + \frac{1}{n^2} \log\left( \frac{1}{2} \right)} \nonumber \\
& {- \frac{(2+\gamma)}{n} \log( n+1) -  \frac{2}{n^2} \log \left( n{\bold{d}(n)} \right)} {- \frac{2( \Xi n + 1)}{n^2} \log n  }.
\end{align}
$\gamma >0$ is some universal constant.
}
\end{thm}
\begin{IEEEproof}
We begin with the achievability claim. We want to construct a code $(f, \varphi)$ {where $f$ maps a graph from $\mathcal{S}$ to a finite set and $\varphi$ maps the range to a reconstructed graph. The result of the encoder $f$ and decoder $\varphi$ on a source graph $\m{G} \in \mathcal{S}$ is $g( \m{G} ) \equiv \varphi( f( \m{G} ))$. We show that there exists a mapping $f$ with bounded range, regardless of the distribution over the set $\mathcal{S}$. This provides us with an upper bound on the combinatorial rate-distortion function.}

We {begin by considering} $\mathcal{S}$ as a disjoint union of edge-types $\rvec{T}(\rvecr{r}, \rvecr{c}, \m{W})$, all with density $\bold{d}(n)$, {so as to obtain a more generalized result for the upper bound. We then restrict $\mathcal{S}$ to a single edge-type class to fit with the constructed lower bound}. For each {edge} type {included in} $\mathcal{S}$ we apply the Covering Lemma (Lemma \ref{lem:coveringLemma}) and construct the reconstruction set as a union of the coverings of each such edge-type. {The Covering Lemma ensures the existence of a reconstruction set that complies with the distortion:}
\begin{align}
\big| B \big| & \leq \sum_{\rvec{T}( \rvecr{r}, \rvecr{c}, \m{W}) \textrm{ s.t. } \rvec{T}( \rvecr{r}, \rvecr{c}, \m{W}) \subseteq \mathcal{S} } \big| B_{\rvec{T}(\rvecr{r}, \rvecr{c}, \m{W})} \big|
\end{align}
where $B_{\rvec{T}( \rvecr{r}, \rvecr{c}, \m{W})}$ denotes the reconstruction set {for the edge-type class} $\rvec{T}( \rvecr{r}, \rvecr{c}, \m{W})$.
Using the bound on the number of possible edge-types as given in \eqref{eq:numberOfPossibleEdgeTypes_assumedDensity}
\begin{align}
 \big| B \big| & \leq e^{ 2n \log( {\bold{d}(n)} +1 ) } \max_{\rvec{T}( \rvecr{r}, \rvecr{c}, \m{W}) \textrm{ s.t. } \rvec{T}( \rvecr{r}, \rvecr{c}, \m{W}) \subseteq \mathcal{S} } \big| B_{\rvec{T}(\rvecr{r}, \rvecr{c}, \m{W})} \big| \nonumber \\
\frac{1}{n^2} \log \big| B \big| & \leq \frac{2}{n} \log( {\bold{d}(n)} +1 ) {+} \max_{\rvec{T}( \rvecr{r}, \rvecr{c}, \m{W}) \textrm{ s.t. } \rvec{T}( \rvecr{r}, \rvecr{c}, \m{W}) \subseteq \mathcal{S} } \frac{1}{n^2} \log \big| B_{\rvec{T}(\rvecr{r}, \rvecr{c}, \m{W})} \big|
\end{align}
{where in the second line we simply took the logarithm and normalized by $n^2$. }
Using Lemma \ref{lem:coveringLemma} {with the assumption on the density \eqref{eq:assumptionFixedDensity_rateDistortion}} we have that
\begin{align}
\frac{1}{n^2} \log \big| B \big| & \leq \max_{ \rvec{T}( \rvecr{r}, \rvecr{c}, \m{W}) \textrm{ s.t. } \rvec{T}( \rvecr{r}, \rvecr{c}, \m{W}) \subseteq \mathcal{S}  } \max_{(\rvecr{d}_r, \rvecr{d}_c) \in \Omega(\Xi)}  \frac{1}{n^2} \left\{ H( \m{F}_{\rvecr{r} \pm \rvecr{d}_r, \rvecr{c} \pm \rvecr{d}_c , \m{W}}  ) -H( \m{F}_{\rvecr{d}_r, \rvecr{d}_c,\m{W}} ) \right\} \nonumber \\
& + \frac{ (2 \Xi {n} + 2)}{n^2} \log n + H_b( \delta) + {\frac{1}{n^2} \log\left( n \bold{d}(n) \right)} + \frac{\gamma}{{n}} \log n + \frac{1}{n} {+} \frac{2}{n} \log( {\bold{d}(n)} +1 ).
\end{align}
Note that as we have used Lemma \ref{lem:coveringLemma} we have that $d( \m{G}, B) \leq \Xi + \frac{\delta}{n}$ for every $\m{G} \in \mathcal{S}$. Now we simply set $f$ to the mapping that gives us $d( \m{G}, f(\m{G})) = d( \m{G}, B)$ and set $\varphi$ to be the identity mapping. This concludes the achievability proof. Note that this construction guarantees with probability one that we comply with the distortion constraint, {regardless of the distribution over the set $\mathcal{S}$. Hence this is a universal bound and we have that
\begin{align}
R_n\left(\Xi + \frac{\delta}{n}\right) & \leq \max_{ \rvec{T}( \rvecr{r}, \rvecr{c}, \m{W}) \textrm{ s.t. } \rvec{T}( \rvecr{r}, \rvecr{c}, \m{W}) \subseteq \mathcal{S}  } \max_{(\rvecr{d}_r, \rvecr{d}_c) \in \Omega(\Xi)}  \frac{1}{n^2} \left\{ H( \m{F}_{\rvecr{r} \pm \rvecr{d}_r, \rvecr{c} \pm \rvecr{d}_c , \m{W}}  ) -H( \m{F}_{\rvecr{d}_r, \rvecr{d}_c,\m{W}} ) \right\} \nonumber \\
& + \frac{ (2 \Xi {n} + 2)}{n^2} \log n + H_b( \delta) + {\frac{1}{n^2} \log\left( n \bold{d}(n) \right)} + \frac{\gamma}{{n}} \log n + \frac{1}{n} {+} \frac{2}{n} \log( {\bold{d}(n)} +1 ).
\end{align} }

{When $\mathcal{S}$ is a single edge-type the above reduces to:
\begin{align}
R_n\left(\Xi + \frac{\delta}{n}\right) & \leq \max_{(\rvecr{d}_r, \rvecr{d}_c) \in \Omega(\Xi)}  \frac{1}{n^2} \left\{ H( \m{F}_{\rvecr{r} \pm \rvecr{d}_r, \rvecr{c} \pm \rvecr{d}_c , \m{W}}  ) -H( \m{F}_{\rvecr{d}_r, \rvecr{d}_c,\m{W}} ) \right\} \nonumber \\
& + \frac{ (2 \Xi {n} + 2)}{n^2} \log n + H_b( \delta) + {\frac{1}{n^2} \log\left( n \bold{d}(n) \right)} + \frac{\gamma}{{n}} \log n + \frac{1}{n}.
\end{align}}

\underline{We now turn to the converse proof.} {We begin this proof by tackling the probabilistic graph rate-distortion, defined in \ref{dfn:ProbabilisticRateDistortion}. In this definition we assume a random graph $\m{F}\sim\{p_{ij}\}$. For the considered distortion measure \eqref{eq:distortion_edge_type}, some $\Xi > 0$ and a small $\delta >0$ the probabilistic graph rate-distortion is given as follows:
\begin{align}\label{eq:prob_rd_UsedinProof}
  & R_n^{\m{F}}\left(\Xi + \frac{\delta}{n},\eps\right) \triangleq \min_{\mathfrak{F}} \frac{\log |\mathfrak{F}|}{n^2}\nonumber \\
  & \text{s.t.}\qquad  \Pr\left(\min_{\m{H}\in \mathfrak{F}} d(\m{F}, \m{H}) > \Xi + \frac{\delta}{n} \right) \leq  \eps.
\end{align}
This is known as the $\epsilon$-fidelity criterion.}
{The converse proof for the probabilistic rate-distortion begins by} assuming that {for any $\epsilon \in (0,1)$} there exists a mapping, $g( \cdot)$, {from any graph in the support of $\m{F}$ to a reconstruction graph} that complies with the $\epsilon$-fidelity criterion for the {considered} random graph {$\m{F}$.} {We obtain a lower bound on the cardinality of the range of $g(\cdot)$, meaning the reconstruction set.}

{Once we obtain such a lower bound we use it to conclude a lower bound on the combinatorial graph rate-distortion $R_n\left(\Xi + \frac{\delta}{n} \right)$.}

We follow the proof given in \cite[Theorem 2.2.3]{csiszar_korner} with slight modifications.
{We consider an arbitrary random graph $\m{F} \in \mathcal{P}(\vec{r},\vec{c}, \m{W}) \cap \mathbb{D}(n \m{W})$. Note that the random graph is any random graph from the corresponding union.}
Starting from the $\epsilon$-fidelity criterion we have that
\begin{align}
Pr \left\{ d( \m{F}, g(\m{F})) \leq \Xi { + \frac{\delta}{n}} \right\} \geq 1- \epsilon 
\end{align}
{We denote the following set}
\begin{align}
\mathcal{T} \equiv \left\{ \m{G} : d( \m{G}, g(\m{G})) \leq \Xi {+ \frac{\delta}{n}}, {\m{G} \textrm{ is in the support of } \m{F}} \right\}. 
\end{align}
{Note that since we now consider an arbitrary random graph from the polytop $\mathcal{P}(\vec{r}, \vec{c}, \m{W})$ we have, according to Lemma \ref{lem:extensionLemma2.12_CK}, that the graphs $\m{G}$ in the above set will belong to $\rvec{T}_{\delta}(\vec{r},\vec{c}, \m{W})$ with high probability.}
Since we further assume that $\m{F} \in \mathbb{D}(n, \m{W})$ we can apply Lemma \ref{lem:lowerBoundedCardinalityOfAnySetOfHighProbability} with $\eta = 1 - \epsilon$ and we have that
\begin{align} \label{eq:thmRateDistortion:lowerBound}
\left| \mathcal{T} \right| \geq e^{ {H( \m{F} ) - n^2 H_b(\hat{\delta}) + \log\left( \frac{1- \epsilon}{2} \right) - n(2+\gamma) \log( n+1) -  \log \left( n{\bold{d}(n)} \right)}}
\end{align}
for some $\hat{\delta} \in {[0, \frac{1}{2})}$ {and the universal constant $\gamma >0$. Note that our assumption in the converse proof is that for any $\epsilon \in (0,1)$ there exists a mapping $g(\cdot)$ that provides $\epsilon$-fidelity. The requirement on the value of $\eta = 1 - \epsilon$ in Lemma \ref{lem:lowerBoundedCardinalityOfAnySetOfHighProbability} is then achieved, at least for a region of values $\epsilon \in (0, M)$ due to our assumption on $n$ and $\bold{d}(n)$ in \eqref{eq:assumptionNandD4usingCardinalityOfHighProbabilitySet_rateDistrotion}, which guarantees the existence of such an $M$. This suffices for our needs, since we eventually take $\epsilon \to 0$.}


Now, denoting by $\mathcal{C}$ {the range of $g(\cdot)$, that is, the set of reconstruction graphs} $\m{H}$ for which there exists some $\m{G}$ {in the support of $\m{F}$,} such that $\m{H} = g(\m{G})$ we have that
\begin{align}
\left| {\mathcal{T}} \right| & \leq \sum_{\m{H} \in \mathcal{C}} \left| \left\{ \m{G}:  d( \m{G}, \m{H} ) \leq \Xi {+\frac{\delta}{n}}, {\m{G} \textrm{ is in the support of } \m{F}} \right\} \right| \nonumber \\
& \leq \left| \mathcal{C} \right| \max_{\m{H} \in \mathcal{C} } \left| \left\{ \m{G}:  d( \m{G}, \m{H} ) \leq \Xi {+\frac{\delta}{n}}, {\m{G} \textrm{ is in the support of } \m{F}} \right\} \right| \nonumber \\
& \leq \left| \mathcal{C} \right| \max_{\m{H} \in \mathcal{C} } \left| \left\{ \m{G}:  d( \m{G}, \m{H} ) \leq \Xi {+\frac{\delta}{n}} \right\} \right| \nonumber \\
& = \left| \mathcal{C} \right| \max_{\m{H} \in \mathcal{C} } \cup_{(\rvecr{d}_r, \rvecr{d}_c) \in \Omega( \Xi), } {\rvec{T}_{\delta} ( \rvecr{d}_r, \rvecr{d}_c | \m{H}, \m{W}) } \nonumber \\
& \leq \left| \mathcal{C} \right| \sum_{(\rvecr{d}_r, \rvecr{d}_c) \in \Omega( \Xi)} \left| \rvec{T}_{{\delta}}( \rvecr{d}_r, \rvecr{d}_c, \m{W}) \right|
\end{align}
{where the equality is due to the fact that the set of all graphs $\m{G}$ that were distorted from graph $\m{H}$ by at most $\Xi + \frac{\delta}{n}$ can be written as a union of conditional $\delta$ edge-type class with respect to the graph $\m{H}$ with parameters that are taken from the set $\Omega(\Xi)$. The additional perturbation of $\delta$ is responsible for the possible additional distortion. Finally, the inequality is due to the fact that the union of conditional $\delta$ edge-types may have some overlaps due to these perturbations. }

We can now simply use the bound on the cardinality of $\Omega( \Xi)$ \eqref{eq:boundOnOmega} and the cardinality of $\rvec{T}_{{\delta}}( \rvecr{d}_r, \rvecr{d}_c, \m{W})$, in {Lemma} \ref{lem:strongTypicalityCardinality}, to determine an upper bound on the above:
\begin{align} \label{eq:thmRateDistortion:upperBound}
\left| {\mathcal{T}} \right| & \leq \left| \mathcal{C} \right| \sum_{(\rvecr{d}_r, \rvecr{d}_c) \in \Omega( \Xi)} \left| \rvec{T}_{{\delta}}( \rvecr{d}_r, \rvecr{d}_c, \m{W}) \right| \nonumber \\
& \leq \left| \mathcal{C} \right| { e^{2( \Xi n + 1) \log n}} e^{ \max_{(\rvecr{d}_r, \rvecr{d}_c) \in \Omega( \Xi)} H\left( \m{F}_{ \rvec{T}( \rvecr{d}_r, \rvecr{d}_c, \m{W})} \right) {+ n^2 H_b(\delta) + \log n + \log \bold{d}(n)}} \nonumber \\
& = \left| \mathcal{C} \right| e^{ {2( \Xi n + 1) \log n + n^2 H_b(\delta) + \log n + \log \bold{d}(n)} +  \max_{(\rvecr{d}_r, \rvecr{d}_c) \in \Omega( \Xi)} H\left( \m{F}_{ \rvec{T}( \rvecr{d}_r, \rvecr{d}_c, \m{W})} \right) }
\end{align}
where in the second inequality we also used the assumption on the changes to the density due to the distortion \eqref{eq:assumptionFixedDensity_rateDistortion} to conclude that the density of the distortion itself can also be bounded from above by the density of the edge-type class.

Putting together \eqref{eq:thmRateDistortion:lowerBound} and \eqref{eq:thmRateDistortion:upperBound} we have that
\begin{align}
e^{ {H( \m{F} ) - n^2 H_b(\hat{\delta}) + \log\left( \frac{1- \epsilon}{2} \right) - n(2+\gamma) \log( n+1) -  \log \left( n{\bold{d}(n)} \right)}} \leq \left| \mathcal{C} \right| e^{ {2( \Xi n + 1) \log n + n^2 H_b(\delta) + \log n + \log \bold{d}(n)} +  \max_{(\rvecr{d}_r, \rvecr{d}_c) \in \Omega( \Xi)} H\left( \m{F}_{ \rvec{T}( \rvecr{d}_r, \rvecr{d}_c, \m{W})} \right) }
\end{align}
meaning that
\begin{align}
\left| \mathcal{C} \right| & \geq e^{ {H( \m{F} ) - n^2 H_b(\hat{\delta}) + \log\left( \frac{1- \epsilon}{2} \right) - n(2+\gamma) \log( n+1) -  \log \left( n{\bold{d}(n)} \right)}} e^{ {- 2( \Xi n + 1) \log n - n^2 H_b(\delta) - \log n - \log \bold{d}(n)} -  \max_{(\rvecr{d}_r, \rvecr{d}_c) \in \Omega( \Xi)} H\left( \m{F}_{ \rvec{T}( \rvecr{d}_r, \rvecr{d}_c, \m{W})} \right) } \nonumber \\
& = e^{ \min_{(\rvecr{d}_r, \rvecr{d}_c) \in \Omega( \Xi)} \left\{ {H( \m{F} )} - H\left( \m{F}_{ \rvec{T}( \rvecr{d}_r, \rvecr{d}_c, \m{W})} \right) \right\} {- n^2\left(H_b(\hat{\delta}) + H_b(\delta) \right) + \log\left( \frac{1- \epsilon}{2} \right) - n(2+\gamma) \log( n+1) -  2 \log \left( n{\bold{d}(n)} \right)} {- 2( \Xi n + 1) \log n  } }.
\end{align}


{The above lower bound is a lower bound on the probabilistic rate-distortion function, assuming some arbitrary $\m{F} \in \mathcal{P}(\vec{r}, \vec{c}, \m{W})$, for any $\epsilon \in (0,1)$:
\begin{align}
R_n^{\m{F}}\left( \Xi + \frac{\delta}{n}, \epsilon \right) \geq & \frac{1}{n^2} \min_{(\rvecr{d}_r, \rvecr{d}_c) \in \Omega( \Xi)} \left\{ {H( \m{F} )} - H\left( \m{F}_{ \rvec{T}( \rvecr{d}_r, \rvecr{d}_c, \m{W})} \right) \right\} {- \left(H_b(\hat{\delta}) +H_b(\delta) \right) + \frac{1}{n^2} \log\left( \frac{1- \epsilon}{2} \right) } \nonumber \\
& - {\frac{(2+\gamma)}{n} \log( n+1) -  \frac{2}{n^2} \log \left( n{\bold{d}(n)} \right)} {- \frac{2( \Xi n + 1)}{n^2} \log n  }.
\end{align}
Recall \eqref{eq:connectionCombinatorialProbabilisticRateDistortion}, that shows how the probabilities rate-distortion function can be used to obtain a lower bound on the combinatorial rate-distortion function. Specifically we further lower bound this by taking the maximum over a subset of random graphs, only those in the $\mathcal{P}(\vec{r}, \vec{c}, \m{W}) \cap \mathbb{D}(n, \m{W})$:
\begin{align}
R_n\left(\Xi + \frac{\delta}{n}\right) \geq \max_{\m{F}} \lim_{\epsilon \to 0} R_n^{\m{F}}\left(\Xi + \frac{\delta}{n},\epsilon\right) \geq \max_{\m{F} \in \mathcal{P}(\vec{r}, \vec{c},\m{W})} \lim_{\epsilon \to 0} R_n^{\m{F}}\left(\Xi + \frac{\delta}{n},\epsilon\right).
\end{align}
We thus obtain the following lower bound on the combinatorial ($0$-fidelity) rate-distortion function by maximizing over all $\m{F} \in \mathcal{P}(\vec{r}, \vec{c}, \m{W})\cap \mathbb{D}(n, \m{W})$:
\begin{align}
R_n\left( \Xi + \frac{\delta}{n} \right) \geq &  \max_{\m{F} \in \mathcal{P}(\vec{r}, \vec{c}, \m{W})\cap \mathbb{D}(n, \m{W})} \lim_{\epsilon \to 0}  R_n^{\m{F}}\left( \Xi + \frac{\delta}{n}, \epsilon \right) \nonumber \\
\geq & \max_{\m{F} \in \mathcal{P}(\vec{r}, \vec{c}, \m{W})\cap \mathbb{D}(n, \m{W})} \frac{1}{n^2} \min_{(\rvecr{d}_r, \rvecr{d}_c) \in \Omega( \Xi)} \left\{ {H( \m{F} )} - H\left( \m{F}_{ \rvec{T}( \rvecr{d}_r, \rvecr{d}_c, \m{W})} \right) \right\} {- \left(H_b(\hat{\delta}) + H_b(\delta)\right) + \frac{1}{n^2} \log\left( \frac{1}{2} \right)} \nonumber \\
& {- \frac{(2+\gamma)}{n} \log( n+1) -  \frac{2}{n^2} \log \left( n{\bold{d}(n)} \right)} {- \frac{2( \Xi n + 1)}{n^2} \log n  } \nonumber \\
= & \frac{1}{n^2} \min_{(\rvecr{d}_r, \rvecr{d}_c) \in \Omega( \Xi)} \left\{ {H( \m{F}_{\rvec{T}(\vec{r}, \vec{c}, \m{W})} )} - H\left( \m{F}_{ \rvec{T}( \rvecr{d}_r, \rvecr{d}_c, \m{W})} \right) \right\} {- \left(H_b(\hat{\delta}) + H_b(\delta)\right) + \frac{1}{n^2} \log\left( \frac{1}{2} \right)} \nonumber \\
& {- \frac{(2+\gamma)}{n} \log( n+1) -  \frac{2}{n^2} \log \left( n{\bold{d}(n)} \right)} {- \frac{2( \Xi n + 1)}{n^2} \log n  }
\end{align}
where the last equality is by the definition of the maximum entropy random graph over the polytop (and since $\m{F}_T$ is also in $\mathbb{D}(n, \m{W})$).
This concludes the proof.}
\end{IEEEproof}

\section{Summary} \label{sec:summary}

The work presented here is an extension of the \emph{method of type} to the more elaborated family of \emph{edge types}. The results concerning this family are in the spirit of the results obtained for types in the standard {method of types}, however they are limited to a family of distributions with a specific form. This is a very limiting restriction. {In Theorem \ref{thm:boundsOnProbabilityGeneralDistribution} we partially extended the observation beyond this limited family, however we were able only to provide a lower bound on the probability to obtain graph $\m{G}$. This is a matter for further study.}


The \emph{edge type} extension is then used as a tool to consider the rate-distortion problem of random directed graphs given a distortion measure that preserved the local structure of the graph. We believe such a measure, or similar measures are more fitting in the compression of random directed graphs. For such measures the standard \emph{method of type} fails to provide a solution. The results obtained thus far are a lower and upper bound on the rate-distortion function. {Note that these bounds are universal, and do not assume anything about the source distribution of the random graph.} They have a similar structure (although, not identical). {Moreover,} the lower bound minimized {an} expression over the set of permissable distortions $(\rvecr{d}_r, \rvecr{d}_c)$ whereas the upper bound maximizes {an} expression over this set. The expressions are reminiscent to the mutual information quantity appearing in the standard rate-distortion solution. {However, we currently do not see how they can be written as the equivalent mutual information in our setting.}

A possible solution to the above problem could be a more elaborated definition for a conditional edge-type. Such a definition was given in \cite{bustin2017lossy} where the distortion was split to two {matrices} {graphs} indicating the edges added and the edges removed {(in \cite{bustin2017lossy} matrices were considered and not graphs)}. In some sense, this is a more natural extension of the vector method of types conditional type definition. However, in the framework of edge-types in the two-dimensional case examined here, such a definition creates dependency on the specific graph examined. For more details on this we refer the reader to a preliminary version of the work \cite{bustin2017lossy}.

\appendix

{\subsection{Example of a structure matrix} \label{appendix:exampleStructureMatrix}
We give here an example taken from \cite{BrualdiSurvay} page 190 - 191. Consider the following normalized edge-type class of graphs over 11 vertices ($n = 11$). The outgoing degrees are:
\begin{align}
\vec{r} = \left(10, 10, 9, 7, 6, 6, 5, 5, 2, 2, 1 \right)
\end{align}
and the ingoing degrees are:
\begin{align}
\vec{c} = \left(11, 9, 9, 8, 8, 5, 5, 3, 3, 1, 1 \right).
\end{align}
The structure matrix of the edge-type is as follows:
\begin{align}
\left[
\begin{array}{cccccccccccc}
63 & 52 & 43 & 34 & 26 & 18 & 13 & 8 & 5 & 2 & 1 & 0 \\
53 & 43 & 35 & 27 & 20 & 13 & 9 & 5 & 3 & 1 & 1 & 1 \\
43 & 34 & 27 & 20 & 14 & 8 & 5 & 2 & 1 & 0 & 1 & 2 \\
34 & 26 & 20 & 14 & 9 & 4 & 2 & 0 & 0 & 0 & 2 & 4 \\
27 & 20 & 15 & 10 & 6 & 2 & 1 & 0 & 1 & 2 & 5 & 8 \\
21 & 15 & 11 & 7 & 4 & 1 & 1 & 1 & 3 & 5 & 9 & 13 \\
15 & 10 & 7 & 4 & 2 & 0 & 1 & 2 & 5 & 8 & 13 & 18 \\
10 & 6 & 4 & 2 & 1 & 0 & 2 & 4 & 8 & 12 & 18 & 24 \\
5 & 2 & 1 & 0 & 0 & 0 & 3 & 6 & 11 & 16 & 23 & 30 \\
3 & 1 & 1 & 1 & 2 & 3 & 7 & 11 & 17 & 23 & 31 & 39 \\
1 & 0 & 1 & 2 & 4 & 6 & 11 & 16 & 23 & 30 & 39 & 48 \\
0 & 0 & 2 & 4 & 7 & 10 & 16 & 22 & 30 & 38 & 48 & 58
\end{array} \right].
\end{align}
The zero positions can be partitioned into four maximal rook paths with three ``gaps''. These ``gaps'' are the non-trivial components, namely: $\{ 1,2\} \times \{10, 11\}$, $\{ 5,6 \} \times \{6,7\}$ and $\{ 9,10 \} \times \{2,3\}$.
From this we can conclude that the adjacency matrices of all graphs in the above normalized edge-type have the following structure:
\begin{align}
\left[
\begin{array}{c}
\begin{array}{cc}
\begin{array}{ccccccccc}
1 & 1 & 1 & 1 & 1 & 1 & 1 & 1 & 1 \\
1 & 1 & 1 & 1 & 1 & 1 & 1 & 1 & 1
\end{array} & \mat{A}_3
\end{array} \\
\begin{array}{ccccccccccc}
1 & 1 & 1 & 1 & 1 & 1 & 1 & 1 & 1 & 0 & 0 \\
1 & 1 & 1 & 1 & 1 & 1 & 1 & 0 & 0 & 0 & 0
\end{array} \\
\begin{array}{ccc}
\begin{array}{ccccc}
1 & 1 & 1 & 1 & 1 \\
1 & 1 & 1 & 1 & 1
\end{array} & \mat{A}_2 &
\begin{array}{cccc}
0 & 0 & 0 & 0 \\
0 & 0 & 0 & 0
\end{array}
\end{array} \\
\begin{array}{ccccccccccc}
1 & 1 & 1 & 1 & 1 & 0 & 0 & 0 & 0 & 0 & 0 \\
1 & 1 & 1 & 1 & 1 & 0 & 0 & 0 & 0 & 0 & 0
\end{array} \\
\begin{array}{ccc}
\begin{array}{c}
1 \\ 1
\end{array} & \mat{A}_1 &
\begin{array}{cccccccc}
0 & 0 & 0 & 0 & 0 & 0 & 0 & 0  \\
0 & 0 & 0 & 0 & 0 & 0 & 0 & 0
\end{array}
\end{array}\\
\begin{array}{ccccccccccc}
1 & 0 & 0 & 0 & 0 & 0 & 0 & 0 & 0 & 0 & 0
\end{array}
\end{array}
\right].
\end{align}
The matrices $\mat{A}_1$, $\mat{A}_2$ and $\mat{A}_3$ are the non-trivial components. In this example they are all $2 \times 2$ matrices with row and column sum vectors of $(1,1)$.  }

\subsection{Proof of Theorem \ref{thm:probabilityOfrandomGraphInEdgeType}} \label{appendix:thm:probabilityOfrandomGraphInEdgeType}
\begin{IEEEproof}
The proof follows the lines of the proof of Theorem 8 in \cite{barvinok2010}. {We first assume that $p_{ij} \in (0,1)$ for all $i \in [n]$ and $j \in [n]$, meaning $p_{ij} \neq 0,1$. We relate to these cases at the end of the proof.}
For all $i,j$ for which {$\{\mat{A}_{\m{W}}\}_{ij} =1$} we have
\begin{align}
\Pr \left( {\{\mat{A}_{\m{F}}\}_{i,j}} = \{\mat{A}_{\m{G}}\}_{i,j} \right) &  = {p_{ij}}^{\{\mat{A}_{\m{G}}\}_{i,j}} (1 - {p_{ij}})^{(1 - \{\mat{A}_{\m{G}}\}_{i,j})} \nonumber \\
& = e^{ \{\mat{A}_{\m{G}}\}_{i,j} \ln {p_{ij}} + (1 - \{\mat{A}_{\m{G}}\}_{i,j}) \ln( 1 - {p_{ij}})}.
\end{align}
Thus, for any $\m{G} \in \rvec{T}( \vec{r}, \vec{c}, \m{W})$ we have
\begin{align} \label{eq:generalProbabilityForG}
\Pr \left( \m{F} = \m{G} \right) & = \prod_{i,j: {\{\mat{A}_{\m{W}}\}_{ij}}=1} e^{ \{\mat{A}_{\m{G}}\}_{i,j} \ln {p_{ij}} + (1 - \{\mat{A}_{\m{G}}\}_{i,j}) \ln( 1 - {p_{ij}})} \nonumber \\
& = e^{ \sum_{i,j:{\{\mat{A}_{\m{W}}\}_{ij}}=1} \left[ { \{\mat{A}_{\m{G}}\}_{i,j} \ln {p_{ij}} + (1 - \{\mat{A}_{\m{G}}\}_{i,j}) \ln( 1 - {p_{ij}})} \right] } .
\end{align}
{
Similarly, for $\m{F} = \m{F}_T$ we have
\begin{align}
\Pr \left( \m{F}_T = \m{G} \right) = e^{ \sum_{i,j:{\{\mat{A}_{\m{W}}\}_{ij}}=1} \left[ { \{\mat{A}_{\m{G}}\}_{i,j} \ln (p_T)_{i,j} + (1 - \{\mat{A}_{\m{G}}\}_{i,j}) \ln( 1 - (p_T)_{i,j})} \right] }.
\end{align} }
Using {Theorem \ref{thm:Barvinok4} (Theorem 8 in \cite{barvinok2010})} claiming that for any $\m{G} \in \rvec{T}(\vec{r}, \vec{c}, \m{W})$
\begin{align} \label{eq:proof:maximumEntropyBarvinokObservation}
{\Pr \left( \m{F}_T = \m{G} \right) =} e^{-H( \m{F}_T )},
\end{align}
we have that
{
\begin{align} \label{eq:proof:maximumEntropyBarvinokObservation2}
e^{-H( \m{F}_T )} = e^{ \sum_{i,j:{\{\mat{A}_{\m{W}}\}_{ij}}=1} \left[ { \{\mat{A}_{\m{G}}\}_{i,j} \ln (p_T)_{i,j} + (1 - \{\mat{A}_{\m{G}}\}_{i,j}) \ln( 1 - (p_T)_{i,j})} \right] }.
\end{align}
Using this in \eqref{eq:generalProbabilityForG} we obtain}
\begin{align}
\Pr \left(\m{F} = \m{G} \right) =  e^{-H( \m{F}_T) - \sum_{i,j:{\{\mat{A}_{\m{W}}\}_{ij}}=1} \left[ \{\mat{A}_{\m{G}}\}_{i,j} \ln \frac{{(p_T)_{i,j}}}{{p_{ij}}} + (1 - \{\mat{A}_{\m{G}}\}_{i,j}) \ln \frac{1 - {(p_T)_{i,j}}}{1 - {p_{ij}}} \right] }. \nonumber
\end{align}
To finalize the proof we need to show that
\begin{align} \label{eq:appendix:probabilityFforG_KLdivergence}
& \sum_{i,j:{\{\mat{A}_{\m{W}}\}_{ij}}=1} \left[ \{\mat{A}_{\m{G}}\}_{i,j} \ln \frac{ {(p_T)_{i,j}}}{{p_{ij}}} + (1 - \{\mat{A}_{\m{G}}\}_{i,j}) \ln \frac{ 1 - {(p_T)_{i,j}}}{1 - {p_{ij}} } \right] = \nonumber \\
& \sum_{i,j:{\{\mat{A}_{\m{W}}\}_{ij}}=1} \left[ {(p_T)_{i,j}} \ln \frac{ {(p_T)_{i,j}}}{{p_{ij}}} + (1 - {(p_T)_{i,j}}) \ln \frac{ 1 - {(p_T)_{i,j}}}{1 - {p_{ij}} } \right] = \nonumber \\
& \sum_{i,j:{\{\mat{A}_{\m{W}}\}_{ij}}=1} D( {(p_T)_{i,j}} || {p_{ij}} ).
\end{align}
{Note that for $(i,j)$ for which $(p_T)_{ij} = 1$ or $(p_T)_{ij} = 0$ the above relationship holds. The reason for this is that when $(p_T)_{ij} = 1$ we have that $\{\mat{A}_{\m{G}}\}_{i,j} = 1$ and when $(p_T)_{ij} = 0$ we have that $\{\mat{A}_{\m{G}}\}_{i,j} = 0$. Thus, for the remainder of the proof we can assume that $\mat{A}_{\m{W}}$ does not contain $(i,j)$'s such that $(p_T)_{ij} = 1$ or $(p_T)_{ij} = 0$ with no loss of generality.}\\
Showing \eqref{eq:appendix:probabilityFforG_KLdivergence} is equivalent to showing that
\begin{align}
\sum_{i,j:{\{\mat{A}_{\m{W}}\}_{ij}}=1} \left[ ( {(p_T)_{ij}} - \{\mat{A}_{\m{G}}\}_{i,j}) \ln \frac{ 1 - {(p_T)_{ij}}}{{(p_T)_{ij}}} - ( {(p_T)_{ij}} - \{\mat{A}_{\m{G}}\}_{i,j}) \ln \frac{1 - {p_{ij}}}{{p_{ij}}}  \right] = 0.
\end{align}
From (\ref{eq:proof:maximumEntropyBarvinokObservation}) we can also conclude that
\begin{align}
\sum_{i,j:{\{\mat{A}_{\m{W}}\}_{ij}}=1} ( {(p_T)_{ij}} - \{\mat{A}_{\m{G}}\}_{i,j}) \ln \frac{1 - {(p_T)_{ij}}}{{(p_T)_{ij}}} = 0.
\end{align}
Similarly, since $\m{F} \in \mathbb{D}(n, \m{W})$ we have that
\begin{align}
& \sum_{i,j:{\{\mat{A}_{\m{W}}\}_{ij}}=1} ( {(p_T)_{ij}} - \{\mat{A}_{\m{G}}\}_{i,j} ) \ln \frac{1 - {p_{ij}}}{{p_{ij}}}  = \nonumber \\
& \sum_{i,j:{\{\mat{A}_{\m{W}}\}_{ij}}=1} ( {(p_T)_{ij}} - \{\mat{A}_{\m{G}}\}_{i,j} ) (a_i + b_j) = \nonumber \\
& \sum_{i,j:{\{\mat{A}_{\m{W}}\}_{ij}}=1} {(p_T)_{ij}} a_i - \sum_{i,j:{\{\mat{A}_{\m{W}}\}_{ij}}=1} \{\mat{A}_{\m{G}}\}_{i,j} a_i + \sum_{i,j:{\{\mat{A}_{\m{W}}\}_{ij}}=1} {(p_T)_{ij}} b_j - \sum_{i,j:{\{\mat{A}_{\m{W}}\}_{ij}}=1} \{\mat{A}_{\m{G}}\}_{i,j} b_j = \nonumber \\
& \sum_i \vec{r}(i) a_i - \sum_i \vec{r}(i) a_i + \sum_j \vec{c}(j) b_j - \sum_j \vec{c}(j) b_j = 0.
\end{align}
{This concludes the proof under the assumption that $p_{ij} \in (0,1)$ for all $i \in [n]$ and $j \in [n]$. Now we want to add these two extreme cases but take into account the requirement that $\m{F}_T \ll \m{F}$. As a result of the requirement of absolute continuity of $\m{F}_T$ with respect to $\m{F}$ we have that when $p_{ij} = 1$ also $(p_T)_{ij} = 1$ and similarly for zero. As such these cases reduce back to the cases of $(p_T)_{ij} = 1$ and $(p_T)_{ij} = 0$ mentioned above, in which case \eqref{eq:appendix:probabilityFforG_KLdivergence} holds. }
This concludes the proof.
\end{IEEEproof}

{
\subsection{Proof of Lemma \ref{lem:extengingTheSetD}} \label{appendix:lem:extendingTheSetD}
\begin{IEEEproof}
To show that the claim holds we need only to show that the set $\mathbb{D}(n, \m{W})$ contains random graphs with a positive probability only for a single edge, and its probability can be chosen freely. If such random graphs are contained in $\mathbb{D}(n, \m{W})$ it is trivial to see that we can construct any general random graph.
We construct such a random graph with a non-zero probability for edge $(m,\ell)$ using the following two vectors:
\begin{align}
\vec{a} & = ( \infty, \ldots, \infty, a_m, \infty, \ldots, \infty) \nonumber \\
\vec{b} & = ( \infty, \ldots, \infty, b_{\ell}, \infty, \ldots, \infty).
\end{align}
The resulting random graph will have the following edge probabilities. For $\left\{\mat{A}_{\m{W}} \right\}_{i,j} = 0$ the probabilities are zero. For $\left\{\mat{A}_{\m{W}} \right\}_{i,j} = 1$ the probabilities are:
\begin{align}
p_{ij} = \frac{ e^{-a_i} e^{-b_j}}{1 + e^{-a_i} e^{-b_j}}.
\end{align}
For the above choice of vectors $\vec{a}$ and $\vec{b}$ we have that for all $(i,j) \neq (m, \ell)$ $p_{ij} = 0$ and
\begin{align}
p_{m\ell} = \frac{ e^{-a_m} e^{-b_{\ell}}}{1 + e^{-a_m} e^{-b_{\ell}}}
\end{align}
can be set to any value (0,1] by choosing finite values for $a_m$ and $b_{\ell}$.
The other direction is immediate.
\end{IEEEproof}
}

{
\subsection{Proof of Theorem \ref{thm:boundsOnProbabilityGeneralDistribution}} \label{appendix:thm:boundsOnProbabilityGeneralDistribution}
\begin{IEEEproof}
Following Lemma \ref{lem:extengingTheSetD} we know that there exists a set $\{\m{F}_1, \ldots, \m{F}_K \}$ such that $\m{F}_k  \sim \{ p_{ij}^k\} \in \mathbb{D}(n , \m{W})$ for all $k \in [K]$ and a vector $\vec{\lambda}$ such that
\begin{align}
\m{F} \sim \left\{ \sum_{k =1}^K \vec{\lambda}(k) p_{ij}^k \right\}.
\end{align}
We write the probability as a product over the edges:
\begin{align} \label{eq:thm:boundsOnProbabilityGeneralDistribution:lowerBound}
\Pr \left( \m{F} = \m{G} \right) & = \prod_{i,j:\left\{ \mat{A}_{\m{W}} \right\}_{i,j} = 1} {\Pr}_{\m{F}} \left( \left\{ \mat{A}_{\m{F}} \right\}_{i,j} = \left\{ \mat{A}_{\m{G}} \right\}_{i,j} \right) \nonumber \\
& = e^{ \sum_{i,j:\left\{ \mat{A}_{\m{W}} \right\}_{i,j} = 1} \ln {\Pr}_{\m{F}} \left( \left\{ \mat{A}_{\m{F}} \right\}_{i,j} = \left\{ \mat{A}_{\m{G}} \right\}_{i,j} \right)} \nonumber \\
& = e^{ \sum_{i,j:\left\{ \mat{A}_{\m{W}} \right\}_{i,j} = 1} \ln  \sum_{k=1}^K \vec{\lambda}(k) {\Pr}_{\m{F}_k} \left( \left\{ \mat{A}_{\m{F}_k} \right\}_{i,j} = \left\{ \mat{A}_{\m{G}} \right\}_{i,j} \right)} \nonumber \\
& \geq e^{ \sum_{i,j:\left\{ \mat{A}_{\m{W}} \right\}_{i,j} = 1} \sum_{k=1}^K \vec{\lambda}(k) \ln  {\Pr}_{\m{F}_k} \left( \left\{ \mat{A}_{\m{F}_k} \right\}_{i,j} = \left\{ \mat{A}_{\m{G}} \right\}_{i,j} \right)}
\end{align}
where the inequality is due to Jensen's inequality, the concavity of the logarithmic function and the monotonically increasing property of the exponential. From this point we continue similarly to the proof of Theorem \ref{thm:probabilityOfrandomGraphInEdgeType} by using the following term for the probabilities:
\begin{align}
{\Pr}_{\m{F}_k} \left( \left\{ \mat{A}_{\m{F}_k} \right\}_{i,j} = \left\{ \mat{A}_{\m{G}} \right\}_{i,j} \right) = \left( p_{ij}^k \right)^{\left\{ \mat{A}_{\m{G}} \right\}_{i,j}} \left( 1 - p_{ij}^k \right)^{\left( 1 - \left\{ \mat{A}_{\m{G}} \right\}_{i,j} \right)}
\end{align}
thus we required that $\m{F}_T \ll \m{F}_k$ for each $k \in [K]$. This concludes the derivation of the lower bound.
\end{IEEEproof}}

\subsection{Proof of Theorem \ref{thm:ExtensionSanov}} \label{appendix:thm:ExtensionSanov}
\begin{IEEEproof}
Given any probability $\m{F} \sim {\{ p_{ij} \}} \in \mathbb{D}(n, \m{W})$ we have that
\begin{align}
{\Pr}_{\m{F}} \left( A \right) = \sum_{\m{F}_T \in \bold{F}} {\Pr}_{\m{F}} \left( \m{G} \in \rvec{T}( \m{F}_T, {\m{W}}) \right)
\end{align}
where we use the notation $\rvec{T}( \m{F}_T, {\m{W}} )$ to denote the edge-type class corresponding to $\m{F}_T {\sim \{ (p_T)_{ij} \}}$ {and restriction graph $\m{W}$}. Using Theorem \ref{thm:probabilityOfAnEdgeType} we have that
\begin{align}
{\Pr}_{\m{F}} \left( A \right) & = \sum_{\m{F}_T \in \bold{F}} {\Pr}_{\m{F}} \left( \m{G} \in \rvec{T}( \m{F}_T, {\m{W}}) \right)   \nonumber \\
& \leq \sum_{\m{F}_T \in \bold{F}} e^{- \sum_{i,j: {\{ \mat{A}_{\m{W}} \}_{i,j}=1}} D( {(p_T)_{ij}} || {p_{ij}} ) }
\end{align}
and
\begin{align}
{\Pr}_{\m{F}} \left( A \right) & = \sum_{\m{F}_T \in \bold{F}} {\Pr}_{\m{F}} \left( \m{G} \in \rvec{T}( \m{F}_T) \right)   \nonumber \\
& \geq \sum_{\m{F}_T \in \bold{F}} (n)^{-2\gamma(2n)} e^{- \sum_{i,j:{\{ \mat{A}_{\m{W}} \}_{i,j}=1}} D( {(p_T)_{ij}} || {p_{ij}} ) } \nonumber \\
& = \sum_{\m{F}_T \in \bold{F}} e^{ - 4 \gamma n\log n } e^{- \sum_{i,j:{\{ \mat{A}_{\m{W}} \}_{i,j}=1}} D( {(p_T)_{ij}} || {p_{ij}} ) } \nonumber \\
& \geq e^{ - 4 \gamma n\log n } e^{- \sum_{i,j:{\{ \mat{A}_{\m{W}} \}_{i,j}=1}} D( {({p_T}^{\star})_{i,j}} || {p_{ij}} ) }
\end{align}
where $\m{F}^{\star}_T \sim \left\{ { ({p_T}^{\star})_{i,j}} \right\}$ {is any member of $\bold{F}$}.
Thus, we have that
\begin{align}
e^{ - 4 \gamma n\log n } e^{-\min_{\m{F}_T \in \bold{F} } \sum_{i,j:{\{ \mat{A}_{\m{W}} \}_{i,j}=1}} D( {(p_T)_{ij}} || {p_{ij}} ) } & \leq {\Pr}_{\m{F}} \left( A \right)  \leq \left| \bold{F} \right| e^{-\min_{\m{F}_T \in \bold{F} } \sum_{i,j:{\{ \mat{A}_{\m{W}} \}_{i,j}=1}} D( {(p_T)_{ij}} || {p_{ij}} ) }.
\end{align}
The cardinality of $\bold{F}$ cannot be more than the number of possible edge-types. If we do not assume anything about the density of the relevant edge-types (see Definition \ref{dfn:density}) the bound is: 
\begin{align} \label{eq:numberOfPossibleEdgeTypes}
\left| \bold{F} \right| \leq \left( (n +1)^{n} \right)^2 = e^{ 2n \log( n +1 ) }.
\end{align}
{If we assume that all relevant edge-types have density $\bold{d}(n)$ then the bound is:
\begin{align} \label{eq:numberOfPossibleEdgeTypes_assumedDensity}
\left| \bold{F} \right| \leq \left( ({\bold{d}(n)} +1)^{n} \right)^2 = e^{ 2n \log( {\bold{d}(n)} +1 ) }.
\end{align}}
This gives us the bound in~\eqref{eq:thm:extensionSanov}.
To see the behavior for large $n$ we rewrite the above as follows:
\begin{align}
e^{n^2 \left( \delta_1 -\frac{1}{n^2} \min_{\m{F}_T \in \bold{F} } \sum_{i,j:{\{ \mat{A}_{\m{W}} \}_{i,j}=1}} D( {(p_T)_{ij}} || {p_{ij}} ) \right) } & \leq {\Pr}_{\m{F}} \left( A \right)  \leq e^{ n^2 \left( \delta_2 -\frac{1}{n^2}\min_{\m{F}_T \in \bold{F} } \sum_{i,j:{\{ \mat{A}_{\m{W}} \}_{i,j}=1}} D( {(p_T)_{ij}} || {p_{ij}} ) \right) }
\end{align}
where
\begin{align}
\delta_1 & = - \frac{4 \gamma}{n} \log(n) \nonumber \\
\delta_2 & = \frac{2}{n} \log(n+1) \textrm{ or } \frac{2}{n} \log(\bold{d}(n)+1). 
\end{align}
For large enough $n$ both $\delta_1$ and $\delta_2$ approach zero. {If the condition in \eqref{eq:SanovConditionOnKLdivergence} holds} the dominant part is the summation over approximately $n^2$ values {(excluding the restricted edges)}. Thus, the expression can be approximated as follows:
\begin{align}
{\Pr}_{\m{F}} \left( A \right)  \approx e^{ -\min_{\m{F}_T \in \bold{F} } \sum_{i,j:{\{ \mat{A}_{\m{W}} \}_{i,j}=1}} D( {(p_T)_{ij}} || {p_{ij}} )}.
\end{align}
This concludes the proof.
\end{IEEEproof}

\subsection{Proof of Lemma \ref{lem:lowerBoundedCardinalityOfAnySetOfHighProbability}} \label{appendix:lem:lowerBoundedCardinalityOfAnySetOfHighProbability}
\begin{IEEEproof}
{Given our assumption in \eqref{lem:assumptionOnNwr2eta} on $\eta$ with respect to $n$ and $\bold{d}_T(n)$ and in view of Lemma \ref{lem:extensionLemma2.12_CK}} we can consider the intersection of the set $A$ with $\rvec{T}_{\delta}( \m{F}_T, {\m{W}} )$ to show that
\begin{align} \label{eq:atLeastEta}
{\Pr}_{\m{F}}\left( A \cap \rvec{T}_{\delta}( \m{F}_T ) \right) \geq \frac{\eta}{2}.
\end{align}
{As pointed out in remark \ref{rem:delta_edge_disjoint_union} }{we may} view $\rvec{T}_{\delta}( \m{F}_T, {\m{W}} )$ as a {disjoint} union of {edge-}type {classes} complying with the requirements in Definition \ref{dfn:strongTypicalityEdgeType}. {We denote these edge-type classes as $\bar{\rvecr{T}}(\bar{\vec{r}}, \bar{\vec{c}}, \m{W})$} (and $\bar{T} = (\bar{\vec{r}}, \bar{\vec{c}}, \m{W})$) and their corresponding {maximum entropy random graphs as}  $\m{F}_{\bar{T}} \sim {\{ \bar{p}_{i,j} \}}$. We have that for each such $\m{F}_{\bar{T}}$
\begin{align}
{\Pr}_{\m{F}}\left( A \cap \rvec{T}( \m{F}_{\bar{T}}, {\m{W}} ) \right) & = \sum_{\m{G} \in  A \cap \rvec{T}(\m{F}_{\bar{T}}, {\m{W}})} {\Pr}_{\m{F}}\left(\m{G}\right)  \nonumber \\
& = \left| A \cap \rvec{T}( \m{F}_{\bar{T}}, {\m{W}}) \right| e^{-H( \m{F}_{\bar{T}}) - \sum_{i,j: {\{\m{A}_{\m{W}}\}_{ij} =1}} D( {\bar{p}_{i,j}} || {p_{i,j}}) } \nonumber \\
& \leq \left| A \cap \rvec{T}( \m{F}_{\bar{T}}, {\m{W}}) \right| e^{-H( \m{F}_{\bar{T}} ) }
\end{align}
where we have used the {result of Theorem \ref{thm:probabilityOfrandomGraphInEdgeType} which states that the probability for all graphs} $\m{G} \in \rvec{T}( \m{F}_{\bar{T}}, {\m{W}} )$ is constant. Now, since the union of edge-types constructing $\rvec{T}_{\delta}( \m{F}_T, {\m{W}})$ is a disjoint union, the probability is the sum of probabilities. Thus we have that
\begin{align}
{\Pr}_{\m{F}}\left( A \cap \rvec{T}_{\delta}(\m{F}_T, {\m{W}} ) \right) & = \sum_{\m{F}_{\bar{T}}} {\Pr}_{\m{F} }\left( A \cap \rvec{T} (\m{F}_{\bar{T}}, {\m{W}} ) \right) \nonumber \\
& \leq \sum_{\m{F}_{\bar{T}}} \left| A \cap \rvec{T}( \m{F}_{\bar{T}}, {\m{W}}) \right| e^{-H( \m{F}_{\bar{T}} ) } \nonumber \\
& \leq e^{2n \log ({\bold{d}_T(n)}+1)} \max_{\m{F}_{\bar{T}}} \left| A \cap \rvec{T}( \m{F}_{\bar{T}}, {\m{W}}) \right| e^{-H( \m{F}_{\bar{T}} ) }
\end{align}
where the maximization is over $\m{F}_{\bar{T}}$ {that correspond to the edge-type classes ${\rvec{T}}(\bar{\vec{r}}, \bar{\vec{c}}, \m{W})$ that comply with Definition \ref{dfn:strongTypicalityEdgeType} with respect to $\rvec{T}(\vec{r}, \vec{c}, \m{W})$, meaning that $\bar{\vec{r}}$ is $\delta$-close to $\vec{r}$ and $\bar{\vec{c}}$ is $\delta$-close to $\vec{c}$. The last inequality also uses \eqref{eq:numberOfPossibleEdgeTypes_assumedDensity} to bound the number of possible maximum entropy random graphs with density $\bold{d}_T(n)$, which is the density of the edge-type classes in $\rvec{T}_{\delta}(\vec{r}, \vec{c}, \m{W})$.}

The above {term is at least} $\frac{\eta}{2}$ according to (\ref{eq:atLeastEta}), thus,
\begin{align}
e^{2n \log ({\bold{d}_T(n)}+1)} \max_{\m{F}_{\bar{T}}} \left| A \cap \rvec{T}( \m{F}_{\bar{T}}, \m{W}) \right| e^{-H( \m{F}_{\bar{T}} ) } \geq \frac{\eta}{2}.
\end{align}
Denoting the maximizing $\m{F}_{\bar{T}}$ as $\m{F}^{\star}$ we have that
\begin{align}
\left| A \cap \rvec{T}( \m{F}^{\star}, {\m{W}}) \right| \geq \frac{\eta}{2} e^{ H( \m{F}^{\star} ) } e^{ -2 n \log({\bold{d}_T(n)} +1)} .
\end{align}
Given the above we have that
\begin{align} \label{eq:boundOnSetA}
| A | & \geq \left| A \cap \rvec{T}( \m{F}^{\star}, {\m{W}}) \right| \nonumber \\
& \geq \frac{\eta}{2} e^{ H( \m{F}^{\star} ) } e^{ -2 n \log( {\bold{d}_T(n)}+1)} \nonumber \\
& = e^{H( \m{F}^{\star} ) + \log\left( \frac{\eta}{2} \right) -2 n \log( {\bold{d}_T(n)}+1)}.
\end{align}
{We want to lower bound $e^{H( \m{F}^{\star} )}$. For this purpose we can look at a general $\m{F}_{\bar{T}}$. }
Using Lemma \ref{lem:strongTypicalityCardinality}, and specifically the upper bound it provides, we have that for every $\bar{T} = ( \bar{\vec{r}}, \bar{\vec{c}}, \m{W})$
\begin{align}
\left| \rvec{T}_{\delta}( \m{F}_{\bar{T}}, \m{W} ) \right| \leq e^{ H( \m{F}_{\bar{T}}) + n^2 H_b( \delta) + \log n {\bold{d}_T(n)}}
\end{align}
{since $\bold{d}_{\bar{T}}(n) = \bold{d}_T(n)$ for all $\bar{T} = ( \bar{\vec{r}}, \bar{\vec{c}}, \m{W})$ complying with Definition \ref{dfn:strongTypicalityEdgeType} with respect to $T = ( {\vec{r}}, {\vec{c}}, \m{W})$.}
{The $\delta$ edge-type class ${\rvec{T}}_{\delta}( \bar{\vec{r}}, \bar{\vec{c}}, \m{W})$ includes the edge-type $\rvec{T}( {\vec{r}}, {\vec{c}}, \m{W})$ as a subset, thus,}
\begin{align}
\left| \rvec{T}_{\delta}( \m{F}_{\bar{T}}, {\m{W}} ) \right| \geq \left| \rvec{T}( \m{F}_{T}, {\m{W}} ) \right|.
\end{align}
Finally, using Theorem \ref{thm:BarvinokThm1} and Lemma \ref{lem:BarvinokDual} we have that
\begin{align}
\left| \rvec{T}( \m{F}_{T}, {\m{W}} ) \right| \geq e^{-\gamma n \log n} e^{H( \m{F}_T )}.
\end{align}
Putting all three observations together we have that for every $\bar{T} = ( \bar{\vec{r}}, \bar{\vec{c}}, \m{W})$
\begin{align}
e^{-\gamma n \log n} e^{H( \m{F}_T )} & \leq e^{ H( \m{F}_{\bar{T}}) + n^2 H_b( \delta) + \log\left( n {\bold{d}_T(n)}\right)} \nonumber \\
e^{H( \m{F}_T )- n^2 H_b( \delta) -\gamma n \log n - \log\left( n{\bold{d}_T(n)} \right) } & \leq e^{ H( \m{F}_{\bar{T}})}.
\end{align}
Using this conclusion on $\m{F}^{\star}$ in \eqref{eq:boundOnSetA} we have that
\begin{align} \label{eq:boundOnSetA_part2}
| A | & \geq e^{H( \m{F}^{\star} ) + \log\left( \frac{\eta}{2} \right) -2 n \log( {\bold{d}_T(n)} +1)} \nonumber \\
& \geq e^{H( \m{F}_T )- n^2 H_b( \delta) -\gamma n \log n + \log\left( \frac{\eta}{2} \right) -2 n \log( {\bold{d}_T(n)} +1) - \log \left( n{\bold{d}_T(n)} \right)}.
\end{align}
This concludes the proof.

\end{IEEEproof}

\subsection{Showing that the distortion constraint holds between $\m{G}$ and all $\m{H} \in \cup_{(\rvecr{d}_r, \rvecr{d}_c) \in \Omega(\Xi)} \rvec{T}_{\delta}(\rvecr{d}_r, \rvecr{d}_c | \m{G}, \m{W})$ in the proof of Lemma \ref{lem:coveringLemma}} \label{appendix:transitionToDeltaEdgeTypeInCoveringLemma}

Consider a graph $\m{G}$. We now show that for any {$\m{H} \in \cup_{(\rvecr{d}_r, \rvecr{d}_c) \in \Omega(\Xi)} \rvec{T}_{\delta}(\rvecr{d}_r, \rvecr{d}_c | \m{G}, \m{W})$}
\begin{align}
{d( \m{G}, \m{H}) \leq \Xi + \frac{\delta}{n}.}
\end{align}
{Let us consider $\m{H}' \in \rvec{T}( \rvecr{d}_r, \rvecr{d}_c | \m{G}, \m{W})$ for some pair $(\rvecr{d}_r, \rvecr{d}_c) \in \Omega( \Xi)$. The distortion \eqref{eq:distortion_edge_type} between $\m{G}$ and any $\m{H}'$ is thus,
\begin{align}
d( \m{G}, \m{H}') & = \frac{1}{n} \max\left\{\|\vec{r}_{\m{G}\oplus \m{H}'}\|_\infty \,,\, \|\vec{c}_{\m{G}\oplus \m{H}'}\|_\infty\right\} \nonumber \\
& = \frac{1}{n} \max\left\{\|\rvecr{d}_r\|_\infty \,,\, \|\rvecr{d}_c\|_\infty\right\} \nonumber \\
& \leq \Xi
\end{align}
where the last inequality is due to the definition of the set $\Omega( \Xi)$.}

{Similarly we can examine $\m{H}' \in \rvec{T}_{\delta}( \rvecr{d}_r, \rvecr{d}_c | \m{G}, \m{W})$ in which case we have:
\begin{align}
d( \m{G}, \m{H}') & = \frac{1}{n} \max\left\{\|\vec{r}_{\m{G}\oplus \m{H}'}\|_\infty \,,\, \|\vec{c}_{\m{G}\oplus \m{H}'}\|_\infty\right\} \nonumber \\
& \leq \frac{1}{n} \max\left\{\|\rvecr{d}_r\|_\infty + \delta \,,\, \|\rvecr{d}_c\|_\infty + \delta \right\} \nonumber \\
& = \frac{1}{n} \max\left\{\|\rvecr{d}_r\|_\infty \,,\, \|\rvecr{d}_c\|_\infty  \right\} + \frac{\delta}{n} \nonumber \\
& \leq \Xi + \frac{\delta}{n}
\end{align}
where the first inequality is due to the definition of conditional $\delta$ edge types, Definition \ref{dfn:strongTypicalityConditional}.}

{As this is valid for any pair $(\rvecr{d}_r, \rvecr{d}_c) \in \Omega(\Xi)$ it is valid for the union of these conditional $\delta$ edge types over all pairs in $\Omega(\Xi)$. }

\subsection{Table of Notations} \label{appendix:TableOfNotations}
\footnotesize{
\begin{center}
\begin{tabular}{ |c|l| }
 \hline
 Notation & Meaning \\
 \hline
 \hline
 $[n]$ & the set $\{1,2,\ldots,n\}$  \\
 $i \gto{G} j$ & a directed edge from $i$ to $j$ exists in the graph $\m{G}$ \\
 $i \ngto{G} j$ & a directed edge from $i$ to $j$ \textbf{does not} exists in the graph $\m{G}$ \\
 $ \{ \mat{A} \}_{ij}$ & the $i,j$ value of matrix $\mat{A}$ \\
 $ \mat{A}[K,L]$ & for $K \subseteq [n]$ and $L \subseteq [n]$ is the sub-matrix of $\mat{A}$ taking the rows in $K$ \\
& and the columns in $L$ \\
 $\mat{A}_{\m{G}}$ & the $n\times n$ binary adjacency matrix equivalently representing the directed graph $\m{G}$ \\
& over the set $[n]$ \\
 $\m{G}\oplus  \m{H}$ & a graph whose adjacency matrix is obtained by a cell-wise XOR of $\mat{A}_{\m{G}}$ and $\mat{A}_{\m{H}}$\\
 $\m{G}\wedge \m{H}$ & a graph whose adjacency matrix is obtained by a cell-wise AND of $\mat{A}_{\m{G}}$ and $\mat{A}_{\m{H}}$\\
$\bar{\m{G}}$ & complimentary graph of $\m{G}$, meaning a graph that contains only and all the edges \\
& $i \ngto{G} j$ \\
 $\m{F}\sim \{p_{ij}\}$ & a random graph in which $i\gto{F} j$ with probability $p_{ij}$ independently over the edges \\
 $\vec{r}_{\m{G}}$ & a vector recording the number of outgoing edges from each of the vertices (Def. \ref{dfn:edgeType}) \\
 $\vec{c}_{\m{G}}$ & a vector recording the number of ingoing edges to each of the vertices (Def. \ref{dfn:edgeType}) \\
 $T_{\m{G}}$ & the \emph{edge-type} of $\m{G}$ which is simply the pair $(\vec{r}_{\m{G}},\vec{c}_{\m{G}})$ (Def. \ref{dfn:edgeType}) \\
 $\rvec{T}_{\m{G}}$ & the edge-type class of the directed graph $\m{G}$ (Def. \ref{dfn:edgeType}) \\
 $\rvec{T}( \vec{r}, \vec{c})$ & the edge-type class directly defined by the pair $( \vec{r}, \vec{c})$ (Def. \ref{dfn:edgeType}) \\
 $T = (\vec{r}, \vec{c}, \m{W})$ or $T = (\m{G}, \m{W})$ & a restricted edge-type (Def. \ref{dfn:RestrictedEdgeType}) \\
 $\rvec{T}( \vec{r}, \vec{c}, \m{W})$ or $\rvec{T}( \m{G}, \m{W})$  & a restricted edge-type class defined by the pair $( \vec{r}, \vec{c})$ and $\m{W}$, \\
& or directly by the graphs $\m{G}$ and $\m{W}$ (Def. \ref{dfn:RestrictedEdgeType}) \\
 $\vec{a}^{\downarrow}$ & for a real-values vector $\vec{a}$, it is a vector with the same components, but sorted \\
&  in descending order \\
 $\mathbb{T}_{\m{G}}$ & Structure matrix for directed graph $\m{G}$. \\
 $\mathbb{T}_{\vec{r}, \vec{c}}$ & Structure matrix for the edge-type class $\rvec{T}( \vec{r}, \vec{c})$ \\
 $ N_1( \cdot)$ & a function receiving a zero-one matrix (an adjacency matrix) and returning the \\
& number of ones \\
 $ N_0( \cdot)$ & a function receiving a zero-one matrix (an adjacency matrix) and returning the \\
& number of zeros \\
$\bold{d}_{\m{G}}(n)$ & the degree-density of the graph $\m{G}$ given as a function on $n$ (Def. \ref{dfn:density}) \\
$\bold{d}_{T}(n)$ & the degree-density of the edge-type $T$ given as a function on $n$ (Def. \ref{dfn:density}) \\
$ \m{F} \ll \hat{\m{F}}$ & the random graph $\m{F}$ is absolutely continuous with respect to the random graph $\hat{\m{F}}$, \\ & meaning each the absolute continuity is held for every $(i,j)$ \\
$\rvec{T}_{\delta}(\vec{r}, \vec{c}, \m{W})$ or $\rvec{T}_{\delta}(\m{F}_T, \m{W})$ & the $\delta$ edge type (Def. \ref{dfn:strongTypicalityEdgeType}) \\
$\rvec{T}( \m{H} | \m{G})$ or $\rvec{T}(\vec{r}, \vec{c} | \m{G})$ & a conditional edge-type class (Def. \ref{dfn:conditionalEdgeType}) \\
$\rvec{T}( \m{H} | \m{G}, \m{W})$ or $\rvec{T}(\vec{r}, \vec{c} | \m{G}, \m{W})$ & a restricted conditional edge-type class (Remark  \ref{rem:restrictedConditionalEdgeType}) \\
$\rvec{T}_{\delta}(\m{H} | \m{G}, \m{W})$ or $\rvec{T}_{\delta}(\vec{r}, \vec{c} | \m{G}, \m{W})$ & a conditional $\delta$ edge-type class (Def. \ref{dfn:strongTypicalityConditional})\\
or $\rvec{T}_{\delta}(\m{F}_T | \m{G}, \m{W})$ & \\
 \hline
\end{tabular}
\end{center}
}
\bibliographystyle{IEEEtran}

\bibliography{ofer_refs_master}

\end{document}